\documentclass[a4paper,12pt]{report}

\catcode`@=11
\def\secteqno{\@addtoreset{equation}{section}%
\def\theequation{\thesection.\arabic{equation}}}
\def\greek2{I\hspace{-.1em}I}
\def\beeqno{\begin{eqnarray*}}
\def\eneqno{\end{eqnarray*}}
\def\beeq{\begin{eqnarray}}
\def\eneq{\end{eqnarray}}
\def\*{e^{i\frac{\theta^{\mu\nu}}{2}\partial_\mu^\alpha \partial_\nu^\beta}}
\def\fra12{\frac{1}{2}}
\def\hata{\hat{A}}
\def\hatlam{\hat{\lambda}}
\def\hatf{\hat{F}}

\def\hatd{\hat{D}}
\def\hsp1{\hspace{1cm}}
\def\cald{{\cal D}}
\def\calf{{\cal F}}
\def\call{{\cal L}}
\def\calo{{\cal O}}

\def\a{\alpha}
\def\b{\beta}
\def\d{\delta}
\def\g{\gamma}
\def\l{\lambda}
\def\L{\Lambda}

\def\O{\Omega}
\def\r{\rho}
\def\s{\sigma}
\def\t{\theta}
\def\e{\eta}
\def\p{\pi}

\def\part{\partial}
\def\tr{{\rm tr}}
\def\trp{{\rm trP}}
\def\pial{2 \pi \alpha'}
\def\hep{hep-th/}
\def\nucl{Nucl. Phys. }
\def\phys{Phys. Rev. }
\def\revl{Phys. Rev. Lett. }
\def\physl{Phys. Lett. }
\def\math{J. Math. Phys.}
\def\mod{Int. J. Mod. Phys. }
\def\modlet{Mod. Phys. Lett. }
\def\prog{Prog. Theor. Phys.}

\def\jhep{JHEP}
\catcode`@=12
\secteqno

\begin{document}

\begin{titlepage}

\vspace{3cm}
\begin{flushright}
CHIBA-EP-126
\end{flushright}

\begin{center}
{\huge
Relations between Non-Commutative  \\
and Commutative Spacetime}\footnote{The review article is a part of the master thesis 
submitted to Chiba University}\\

\vspace{1cm}

Ken-Ichi Tezuka\footnote{e-mail: tezuka@physics.s.chiba-u.ac.jp}  \\

\vspace{1cm}

Graduate School of Science and Technology Chiba University, Japan \\

\vspace{5cm}

{\large  Abstract}   \\
\end{center}
Spacetime non-commutativity appears in string theory. In this paper, the 
non-commutativity in string theory is reviewed. At first we review 
that a Dp-brane is equivalent to  a configuration of infinitely many D($p-2$)-branes. 
If we consider the worldvolume as that of the Dp-brane, coordinates of the Dp-brane is 
commutative. On the other hand if  we deal with the worldvolume as that of the 
D($p-2$)-branes, since coordinates of many D-branes are promoted to matrices the 
worldvolume theory is non-commutative one. Next we see that using a point splitting 
reguralization gives a non-commutative D-brane, and a non-commutative gauge field can 
be rewritten in terms of an ordinary gauge field. The transformation is called the 
Seiberg-Witten map. And we introduce second class constraints as
boundary conditions of an open string. Since Neumann and Dirichlet 
boundary conditions are mixed in the constraints when the open string 
is coupled to a NS B field, the end points of the open string is 
non-commutative. 
\end{titlepage}

\tableofcontents
\newpage
\begin{center}
{\large Acknowledgments}
\end{center}
There are some people who have helped me to write the master thesis. 
\par
I'm most grateful to my advisor Tadahiko Kimura for giving me appropriate advises 
when my study in the string theory did not come off well, and reading the manuscript 
of the master thesis. 
\par
I would like to thank Nobuyuki Ishibashi (KEK) for giving clear talk at Chiba 
University. The contents in chapter 3 is based on his talk. I'm also thankful to 
Koichi Murakami (Osaka Univ.) for giving a pedagogical lecture at Tateyama, Chiba. 
By his talk I have my understanding in the string theory increased. 
\par
I'm obliged to Shuuichiro Mori for studying string theory and field theory for two 
years with me. The discussion with him is helpful. 
I also thank Atsushi Nakamura for 
informing me of important papers for gauge fixing in the string theory. I would like 
to thank Seiji fukuzawa for pointing out misprints in the manuscript.

\chapter{Introduction}
It goes without saying that the common and traditional language in particle 
physics is the quantum field theory\cite{weinberg}. The theory represents the 
interaction, namely electroweak and strong, among {\it elementary} particles and 
matter (electrons and quarks etc.) by fields. Quantum field theory is used also in  
cosmology, solid state physics and so on. When we deal with phenomena with 
respect to elementary particles, gravitation can be ignored since this is weaker 
than other forces  in almost cases. On the other hand, gravitation is the 
most conspicuous force in the macroscopic world. Gravitation is successfully 
described by general relativity \cite{foster} which is also able to be 
viewed as the 
theory of spacetime. In the macroscopic world the quantum effect can be negligible. 
In other words the macroscopic world is classical which is a state with large 
quantum number. The universe should be described by the general relativity and it is 
adequate to be treated classically. Since the early universe, however, is 
microscopic, we also need to consider quantum effects to it. A relation between the 
metric of spacetime and matter is given by Einstein equation which is 
\beeqno
R_{\mu \nu } -\frac{1}{2}g_{\mu \nu} R=8 \p GT_{\mu \nu}.
\eneqno
The left hand side describes geometric properties of spacetime, and the right hand 
side describes matter fields. We can deal with the matter fields in the framework of 
quantum field theory. The equation means that quantum effects of the matter fields 
affects the spacetime. Hence we need {\it quantum gravity}. However it is difficult 
to deal with the quantum aspects of gravitational force in terms of quantum field 
theory. 
\par
At present there are mainly two types of standpoints for quantum gravity 
\cite{horowitz}. One of which is that we do not change the classical theory of 
gravity, and we quantize it. This is called quantum geometry. Another one is that we 
should construct a new classical theory of gravity which is reduced to the general 
relativity in a limit and quantize it. The most typical one is superstring 
theory \cite{green,polchinskistring,polchinskitasi,kiritsis,bachas}. Quantization 
of the general relativity is essentially equivalent to that of spacetime. Quantum 
gravity naively identified with non-commutative geometry. On non-commutative 
space, coordinates do not commute with each other; $[X^{\mu},X^{\nu}]\neq 0$. 
Recently it is attracted much attention by a lot of authors that we can regard the 
spacetime where strings live as non-commutative space 
\cite{connesgeo,connes,douglas,seiberg,ishibashirelation,katononcom}. In this master 
thesis we would like to review the subject. 
\par
Here we would like to see necessities of introducing strings. It have been proven 
that gauge theories  except for the general relativity (which is a kind of gauge 
theory) are {\it renormalizable}. The 
renormalization of quantum electrodynamics is formulated independently by Schwinger, 
Feynman and Tomonaga. Tomonaga have however thought that a  renormalization is not 
a final resolution but a temporary method to the problem of divergences. And in 
those days the theory of weak interaction is Fermi theory. This theory is not 
renormalizable. For these reasons many people have made efforts to construct a 
theory which does not have divergences. One of the efforts is non-local field theory. 
But there is 
no one who successes to construct consistent non-local field theory. Up to now 
elementary particles have been regarded as point like objects which is coincide with 
experimental data. Since  divergences come from the point like interactions in 
quantum field theory, some physicists have thought that quantum theory of extended 
objects does not give such divergences. This is one of reasons why we investigate 
string theory.  
\par
Next let us see the strong interaction. In scattering experiments unstable particles 
appear as  resonances. For resonances there is a relation between mass and spin;
\beeq
m^2=\frac{J}{\a'}                           \label{regge}
\eneq
where $\a' \sim 1 (GeV)^{-2}$. This is tested up to $J=\frac{11}{2}$. We can not 
describe this behavior by field theory. Then we need a model to do this. 
\par
Venetiano has suggested a scattering amplitude of hadrons phenomenologically. It 
was suggested by Nambu and Goto that the Venetiano amplitude is derived from the 
bosonic string theory. Strings are one dimensionally extended objects. The strings 
are sorted into open and closed strings. The open string has end points and the 
closed string does not. The theory is defined perturbatively in the sense that 
there is no interaction term in the string action and the interaction is included 
by vertex operators. The relation (\ref{regge}) can be derived from this theory. 
The bosonic string theory has some difficulties. This contains a tachyon which has 
mass $m^2<0$. The presence of the tachyon makes systems unstable since the potential 
is not bounded from the below. In order to preserve the Lorentz invariance at quantum 
level the spacetime dimension should be 1+25\footnote{It was pointed out by Kato and 
Ogawa \cite{kato} that the critical dimension $D=26$ is needed in order to preserve 
the nil-potency of BRS charge. However it was suggested by Abe and Nakanishi 
\cite{abe,takahashi,kawano} that there is the possibility that the critical 
dimension is ill-defined. Their claim is that the fact ``$Q^2=0$ only for $D=26$'' 
is true only if we use the conformal gauge which was used in \cite{kato}. We need 
further investigation for the subject. We would like to discuss this in another 
publication.}. However our world is 1+3 dimension. Although in the real world there 
are a lot of fermions, the bosonic string theory does not contain fermion.     
\par
At the same period, it was suggested by Weinberg and Salam that the weak interaction 
is described by spontaneously broken $SU(2)\times U(1)$ gauge theory which is called 
electroweak theory. The strong interaction is described by $SU(3)$ gauge theory 
which is called quantum chromodynamics (QCD). It was proved by t'Hooft that these 
theories are renormalizable. Because of the success of the gauge theories, studies 
of string theory as a theory of strong interaction have disappeared. 
\par
In the spectrum of closed string theory, there is a spin 2 field. Scherk, Schwarz 
and Yoneya have interpreted it as a graviton. In other words there is the 
possibility that the string theory is not the theory of hadron, but quantum theory 
of gravitation. It is believed that the theory has no UV divergence. 
\par
Neveu, Schwarz and independently Ramond have constructed the dual model which 
contains fermions \cite{neveu,ramond}. The dual model has worldsheet scalars and 
worldsheet fermions. The NSR string theory has the worldsheet supersymmetry. The 
supersymmetry 
is defined as a symmetry which exchanges bosons and fermions. We need to determine 
boundary conditions of strings. For fermionic variables, there are two types of 
conditions; Neveu-Schwarz (NS) and Ramond (R) boundary conditions. Because of the  
existing of two types of conditions, degrees of freedom increase. Hence we have to 
project out the extra degrees of freedom. This is called GSO (Gliozzi, Schwarz and 
Olive) projection \cite{gliozzi}.  After performing the GSO projection the NSR string 
theory has the spacetime supersymmetry, and does not contain a tachyon. In the theory 
the worldsheet supersymmetry is manifest but the spacetime supersymmetry is not.  
\par
Green and Schwarz \cite{greensuper,greencovariant,siegelstring} have constructed the 
manifestly spacetime supersymmetric string theory which is equivalent to the NSR 
string theory. The Green-Schwarz string does not have manifest worldsheet 
supersymmetry and it is difficult to quantize it with covariant gauge fixing 
conditions \cite{hori} because of the $\kappa$ symmetry \cite{siegel} which is needed 
in order to have equivalence with the NSR string. 
\par
The critical dimension of superstring theory is D=10 
\cite{ito,deser,brink,howe,howesuper}. We should solve this problem. The most popular 
method is a compactification method. The word ``compactification'' means that 
some space directions are periodic and their radii are very small ($R \to 0$). The 
compactified directions are so small that we can not find motion of particles along 
to the compactified directions. Let us see a plain example. A stick is a three 
dimensionally extended object. However if the stick is seen from afar, this looks as 
if this is one dimensional object. Of course this is not the only possibility. 
Recently the brane world scenario is studied by a lot of authors \cite{randall}. 
\par
We can construct five kinds of the perturbative string theories which are sorted by 
their symmetries. Type I theory has N=1 supersymmetry where N is the number of the 
supercharges. Type \greek2A (\greek2B) theory has N=2 supersymmetry whose 
chiralities are opposite (same). We can also construct string theories whose 
left mover is bosonic string and right mover is superstring. The only allowed 
gauge groups are $E_8 \times E_8$ and $SO(32)$. These are called $E_8 \times E_8$ 
and $SO(32)$ Heterotic string theories respectively.       
\par
It is expected that the string theory describes physics at the Planck scale. Hence 
we need to construct the low energy effective theory. However we do not know how to 
choose a true way to compactify the extra dimensions. There are infinitely many ways 
to construct the effective theories. This is the limit of applicability of the 
perturbative string theory. It is guessed that the true vacuum is determined by the 
construction of non-perturbative string theory. 
\par
When we see the dynamics of open strings we should determine boundary conditions of 
them. There are two types of conditions; Neumann and Dirichlet boundary conditions. 
For the Dirichlet, an end point of string can not move along to the direction. When 
the number of the 
directions to which open string satisfies the Dirichlet condition is p, we can 
represent this situation by the open string which is attached to a p-dimensional 
object and freely moves on the brane. We call it as Dp-brane.   
\par
The ten dimensional supergravity is the effective theory of the superstring theory 
in the sense that this contains only massless modes, and all of the massive modes 
are integrated out. The theory has solitons. We 
can understand non-perturbative effects through the solitons. The effective theory 
of string theory has a p-dimensional solitons. These solitons correspond to  
D-branes in string theory \cite{polchinskirr}. 
\par
Various non-perturbative definitions of string theory are proposed which use lower 
dimensional D-branes as fundamental degrees of freedom.  In the \greek2B matrix 
model \cite{ishibashimatrix} D-instantons are fundamental degrees of freedom. And 
in the BFSS matrix theory \cite{banks} it is conjectured that M-theory in the 
infinite momentum frame is equivalent to a large N limit of the theory of N 
D0-branes. It is guessed that the string theories correspond to a weak coupling 
limit of the M-theory. But the theory does not have any correct definition yet. In 
these matrix models coordinates are represented by matrices. Because of this the 
spacetime becomes naturally non-commutative. It was pointed out by Connes, Douglas 
and Schwarz \cite{connes} that M-theory with background constant three form tensor 
field compactified on a torus can be identified with matrix theory compactified on 
a non-commutative torus. Corresponding to this, the string theory with background NS 
B field is equivalent to the string theory on a non-commutative space \cite{douglas}.
\par        
The non-commutative spacetime is studied not only in the string theory, but also in 
field theory. Pauli has suggested that quantized spacetime can be used as a 
regulator in field theory. Ydri \cite{ydri} has shown that all infinities in 
$\phi^4$ theory can be removed perturbatively by choosing appropriate 
non-commutative space.
\par
Some authors have shown that non-commutativities come from differences of 
viewpoints. A Dp-brane with a constant gauge field is equivalent to $\infty$ 
D($p-2$)-branes \cite{ishibashibrane}. The worldvolume theory of a Dp-brane is 
ordinary gauge theory, and that of $\infty$ D($p-2$)-branes is non-commutative 
gauge theory. Another way to give non-commutative coordinates is related to   
regularization methods. In this paper, we would like to see mainly relations between 
commutative and non-commutative field theories.
\par   
The subjects covered in this paper are organized as follows. In chapter 2 we see a 
relation between T-duality and D-brane. Non-commutativity in the string theory  
always appears at end points of strings at which D-branes are. The T-duality in 
closed string theory is defined as the exchange of Kaluza-Klein mode and winding 
mode, and simultaneously the exchange of a compactification radius $R$ and 
$\hat{R}\equiv \frac{\a'}{R}$. Since in open string theory there is no winding 
mode, we define it as the exchange of 
Dirichlet condition and Neumann condition. We see the definition of boundary state 
which represents a boundary of a string (D-brane). In chapter 3 we construct the 
boundary state corresponding to infinite D(p-2)-branes. And we show equivalence 
between a Dp-brane and $\infty$ D(p-2)-branes \cite{ishibashibrane}. An important 
point is that the theory of a Dp-brane is the ordinary DBI theory, and that of 
D(p-2)-branes is a gauge theory on non-commutative space. In chapter 4 another 
relation between commutative and non-commutative gauge theories is given. We see 
that products of functions on non-commutative space are Moyal products. If there is 
a constant background NS B field, coordinates of end points of open string become 
non-commutative. We observe this in terms of conformal field theories (CFT). In 
quantum field theories when we calculate an amplitude, its value is infinite. In 
order to make theory converge, we need to regularize the theory. At the quantum 
level symmetry depends on the regularization method. We find that the use of point 
splitting regularization gives non-commutative gauge theory. The Seiberg-Witten 
map which relates gauge theory on ordinary space with non-commutative gauge theory 
is reviewed \cite{seiberg} in chapter 5. In the framework of the operator 
formalism, we can understand that the mixed type (coordinate and canonical 
momentum) boundary condition is a source of non-commutativity when we regard the 
boundary conditions as primary constraints \cite{chunoncom,ardalan,chu,sheikh} 
which is reviewed in chapter 6. The final chapter contains summary and remarks.  


\chapter{T-Duality and D-Brane}
%
%

\section{Open String and Boundary Conditions}
In the conformal gauge, the bosonic string action is stationary  
where the coordinate $X^{\mu}$ satisfies 
the equation of motion:
\[
\partial_{\alpha}\partial^{\alpha}X^{\mu}=0 ,
\]
and furthermore in the open string case the Neumann boundary condition 
\[
\partial_{\sigma} X^{\mu}|_{\sigma =0,\pi}=0 
\]
or the Dirichlet condition
\[
\delta X^{\mu} |_{\sigma =0,\pi}=0.
\]
\par
These boundary conditions are described by a D-brane at which the end points of the 
open string are. Along longitudinal directions of the D-brane, the string satisfies 
the Neumann condition, on 
the other hand along transverse directions of the D-brane the string satisfies the 
Dirichlet condition. The end points of the open string can not leave the D-brane. 
There are four kinds of choices of the boundary 
conditions. We write down the solutions of the Eq. of motion 
for each cases below. \\

\vspace*{0.5cm}
\emph{N-N boundary condition} \\
If an open string satisfies the Neumann boundary condition at both 
end points, the solution of the Eq. of motion is 
\beeq
X^{\mu}=x^{\mu} + 2 \alpha ' p^{\mu} \tau +i \sqrt{ 2 \alpha '}
\sum_{n \neq 0} \left( \frac{\alpha _n ^{\mu}}{n} e^{-i n\tau} \cos n \sigma
\right).                                     \label{2-11}
\eneq

\vspace*{0.5cm}
\emph{D-D boundary condition} \\
The solution is 
\begin{eqnarray*}
X^{\mu}=\frac{c^{\mu}(\pi - \sigma) +d^{\mu} \sigma}{\pi} - \sqrt{ 2 \alpha '}
\sum_{n \neq 0} \left( \frac{\alpha _n ^{\mu}}{n} e^{-i n\tau} \sin n \sigma
\right).
\end{eqnarray*}

\vspace*{0.5cm}
\emph{D-N boundary condition} \\
\begin{eqnarray*}
X^{\mu}=c^{\mu} - \sqrt{ 2 \alpha '}
\sum_{r \in \mathbf{Z} +\frac{1}{2}}
 \left( \frac{\alpha _r ^{\mu}}{r} e^{-ir \tau} 
\sin r \sigma \right)
\end{eqnarray*}

\vspace*{0.5cm}
\emph{N-D boundary condition} \\
\begin{eqnarray*}
X^{\mu}=d^{\mu}  +i \sqrt{ 2 \alpha '}
\sum_{r \in \mathbf{Z} +\frac{1}{2}}
 \left( \frac{\alpha _r ^{\mu}}{r} e^{-i r\tau}
 \cos r \sigma \right)
\end{eqnarray*}
%
%
\section{T-Duality and Closed String}
In this section we see about relations between T-duality and D-branes 
\cite{polchinskicombi,vecchia}. 
At first consider bosonic closed strings with a condition 
$X^{\mu} (\sigma)=X^{\mu} (\sigma + \pi)$. The solution of the equation of motion is 
given by 
\begin{eqnarray*}
X^{\mu} (\tau, \sigma)&=& x^{\mu}+\sqrt{2 \alpha '} 
(\alpha _0 ^{\mu} +\tilde{\alpha} _0 ^{\mu}) \tau -
\sqrt{2 \alpha '} 
(\alpha _0 ^{\mu} -\tilde{\alpha} _0 ^{\mu}) \sigma  \\
&&\hsp1+i \sqrt{\frac{\alpha '}{2}} \sum _{n \neq 0} 
\frac{1}{n} \left(\alpha _n ^{\mu} e^{-2in(\tau -\sigma)}
+\tilde{\alpha} _n ^{\mu} e^{-2in(\tau +\sigma)}
\right).
\end{eqnarray*}
In uncompactified directions, the term proportional to $\sigma$ 
is not allowed. Then $\alpha _0 ^{\mu} -\tilde{\alpha} _0 ^{\mu}
=0$. According to the N\" other method, it is easy to see 
the conserved momentum of the string is
\begin{eqnarray}
p^{\mu}=\frac{1}{\sqrt{2 \alpha '}}(\alpha _0 ^{\mu} 
+\tilde{\alpha} _0 ^{\mu}).  \label{2-2}
\end{eqnarray}
We consider the situation that the 25th direction of spacetime is compactified 
into $S^{1}$. This 
statement is equivalent to the condition in which two points in the compactified 
direction are identified;
\begin{eqnarray}
x^{25} \sim  x^{25} +2 \pi R          \label{2-1},
\end{eqnarray}
where R is the radius of $S^{1}$. The generator of the translation along 
the compactified direction is $e ^{ipx}$. From the identification 
(\ref{2-1}),
it can be understood that physical states are invariant under  
the translation from $x=0$ to $x=2 \pi R$, then the 25th direction of 
the momentum (\ref{2-2}) must be 
\begin{eqnarray}
p^{25}=\frac{n}{R} \ \ \ \ \ \ \ \ \ n \in Z.   \label{2-5}
\end{eqnarray}
In order to satisfy the condition $X^{\mu} (\sigma)
=X^{\mu} (\sigma +\pi)$, the term proportional to $\sigma$ must be
\begin{eqnarray*}
\pi \sqrt{2 \alpha '} (\alpha _0 ^{25} -\tilde{\alpha} _0 ^{25})
=2 \pi wR
\end{eqnarray*}
where w is a winding number of the closed string about the compactified 
direction. From above equations, one find 
\begin{eqnarray*}
\alpha^{25}_0&=& \sqrt{\frac{\alpha'}{2}} \left[  \frac{n}{R}
+\frac{wR}{ \alpha '} \right]   \\
\tilde{\alpha^{25}_0} &=& \sqrt{\frac{\alpha'}{2}} \left[  \frac{n}{R}
-\frac{wR}{ \alpha '} \right]
\end{eqnarray*}
\par
With these situations, we would like to see the mass shell condition. 
To do this the zero mode of the Virasoro generators $L_n, \tilde{L}_n$ 
are needed. These are written as the coefficients of the mode 
expansion of the energy momentum tensor
\begin{eqnarray*}
T_{--}= 4 \alpha '\sum_{n} L_n e^{-2in(\tau -\sigma)}   \\
T_{++}=4 \alpha ' \sum_{n} \tilde{L}_n e^{-2in(\tau +\sigma)}
\end{eqnarray*}
which can be rewritten:
\begin{eqnarray}
L_{n}=\frac{1}{4 \pi \alpha '} \int _0 ^{\pi} d \sigma T_{--}
e^{2in (\tau-\sigma)}                     \label{2-7}               \\
\tilde{L}_{n}=\frac{1}{4 \pi \alpha '} \int _0 ^{\pi} d \sigma T_{++}
e^{2in (\tau+\sigma)}.                       \label{2-8}
\end{eqnarray}
The energy momentum tensor is 
\begin{eqnarray}
T_{++} = \partial _{+} X^{\mu} \partial_{+} X_{\mu}       \label{2-9}          \\
T_{--}=\partial _{-} X^{\mu} \partial _{-} X_{\mu},             \label{2-10}
\end{eqnarray}
which are obtained by a variation of the string action with respect to 
the world sheet metric\cite{green}. Performing the integrations in 
(\ref{2-7}) and (\ref{2-8}) with (\ref{2-9}) and (\ref{2-10}) gives $L_{n}$ and 
$\tilde{L}_n$ which are
\begin{eqnarray*}
L_m =\frac{1}{2}\sum_{-\infty} ^{\infty}
\alpha _{m-n} \cdot \alpha _{n}    \\
\tilde{L}_m =\frac{1}{2}\sum_{-\infty} ^{\infty}
\tilde{\alpha} _{m-n} \cdot \tilde{\alpha} _{n}.
\end{eqnarray*}
The mass shell condition is 
\begin{eqnarray*}
M^2 = \frac{2}{\alpha '} \sum _{n=1} ^{\infty}
(\alpha_{-n} \cdot \alpha_{n} +\tilde{\alpha}_{-n} \cdot 
\tilde{\alpha}_{n} ) - \frac{4}{\alpha '} 
+\left(\frac{n}{R} \right) ^2 +\left(\frac{w R}{\alpha '} 
\right) ^2
\end{eqnarray*}
These equations are invariant under the exchange of the Kaluza$-$Klein 
mode with the winding mode, and simultaneously $R$ with $\hat{R}$:
\begin{eqnarray*}
w   \iff       n     \ \ \ \ \ \ \ ,\ \ \ \ \ \ \
R  \iff    \hat{R} \equiv \frac{\alpha '}{R}.
\end{eqnarray*}
This is T-duality transformation. $\hat{R}$ is a 
comapactification radius of the T-dual theory. In terms of 
zero modes of the string oscillator, this transformation is 
written as
\begin{eqnarray}
\alpha _0  \to  \alpha _0 \ \ \ \ \ \ \ , \ \ \ \ \ \ 
\tilde{\alpha} _0  \to - \tilde{\alpha} _0
                          \label{2-3}
\end{eqnarray}
Under this transformation, the physical space is 
changed. One can notice this fact by seeing the form 
of the Virasoro Operators.
In $\tilde{L}_m$, a term with $\tilde{\alpha_0}$ is contained.
Because we would like not to change the physical space 
under the T-duality transformation, we have to extend 
the definition of the T-duality transformation (\ref{2-3}) 
to the non-zero modes:
\begin{eqnarray}
\alpha _n  \to \alpha _n \ \ \ \ \ \ \ , \ \ \ \ \ \ 
\tilde{\alpha} _n  \to - \tilde{\alpha} _n.
\end{eqnarray}
\par
It is convenient to deal with the T-duality in terms of string target space 
coordinates. The string target space coordinate $X^{\mu}$ can be decomposed 
into the left and right moving modes:
\begin{eqnarray}
X^{\mu} (\tau , \sigma)&=&X^{\mu} _L(\tau - \sigma)
+X^{\mu} _R(\tau + \sigma)    
\end{eqnarray}
where 
\begin{eqnarray*}
X^{\mu} _L(\tau - \sigma)&=& x^{\mu}+\sqrt{2 \alpha '} 
(\tau - \sigma)\alpha _0 ^{\mu} 
+i \sqrt{\frac{\alpha '}{2}} \sum _{n \neq 0} 
\frac{1}{n} \alpha _n ^{\mu} e^{-2in(\tau -\sigma)}   \\
X^{\mu} _R(\tau + \sigma)&=& x^{\mu}+\sqrt{2 \alpha '} 
(\tau + \sigma)\tilde{\alpha} _0 ^{\mu} 
+i \sqrt{\frac{\alpha '}{2}} \sum _{n \neq 0} 
\frac{1}{n} \tilde{\alpha} _n ^{\mu} e^{-2in(\tau +\sigma)} .  
\end{eqnarray*}
We will extend the definition of the T-duality transformation to 
one including  center of mass coordinates $x^{\mu}$;
\begin{eqnarray}
X \to \hat{X}=X_L -X_R   \label{2-4}
\end{eqnarray}
which is called T-dual coordinate.  
In the next section we will show that the T-dual symmetry is a key concept 
to understand D-branes in open string theory.
%
%
\section{Open Strings and T-Duality}
Open string theory does not have a winding mode. From this 
point of view, one may think that there is not the T-duality 
in open string theory. But this idea is inconsistent. 
\par
For example, an open string 1-loop worldsheet (cylinder) diagram can also be viewed 
as a closed string tree diagram. Therefore open string theory contains closed strings. 
Let us consider a  $R \to 0$ limit. In this limit, if we see the worldsheet as the 
open sting diagram, the K-K mode is infinitely massive. Hence the only allowed 
state which does not decouple from the theory is with $n=0$ (i.e. zero 
momentum state). This suggests the open string can not have momentum 
along compactified directions, in other words, be living in 
D-k dimensional space (k is the number of the compactified dimensions). On 
the other hand, in the closed string channel, by virtue of the 
T-duality, the $R \to 0$ limit correspond to the $\hat{R} \to \infty$ 
in the T-dual theory. This makes the closed string possible to have 
momentum along the compactified directions. This is not consistent. 
\par
Then let us define the T-duality in the open string theory so that it will solve this 
inconsistency. It is natural to define the T-dual coordinates also in open string 
theory as (\ref{2-4}) where the left and right movers are 
\begin{eqnarray*}
X_L (\tau - \sigma )=\frac{1}{2}x +\sqrt{2 \alpha '}(\tau - \sigma )\alpha _0 
+i\sqrt{2 \alpha '} \sum_{n \neq 0} \frac{\alpha _n}{n}
e^{-in(\tau - \sigma )}    \\
X_R (\tau + \sigma )=\frac{1}{2}x +\sqrt{2 \alpha '}(\tau + \sigma )\alpha _0 
+i\sqrt{2 \alpha '} \sum_{n \neq 0} \frac{\alpha _n}{n}
e^{-in(\tau + \sigma )}.
\end{eqnarray*}
The T-dual coordinates satisfy the conditions
\begin{eqnarray}
\partial_{\tau} X \to \partial_{\tau} \hat{X}= -\partial_{\sigma}X  
                    \nonumber \\
\partial_{\sigma} X \to \partial_{\sigma} \hat{X}= -\partial_{\tau}X.
\end{eqnarray}
These facts imply that the Neumann and Dirichlet boundary 
conditions are interchanged by T-duality transformation. In terms of the D-brane, 
the D-branes in the original theory and the T-dual theory are at angle 
$\frac{\p}{2}$.
\par
It is easy to see that an open string which satisfies these  
boundary conditions is interpreted to be attached to a hyper-surface which is 
called a Dp-brane, where p is the number of uncompactified 
(non-T-dualized) directions. The distance between two T-dual coordinates 
of endpoints of the open string is 
\begin{eqnarray}  
\hat{X}(\pi)-\hat{X}(0)=-2 \pi n \hat{R}.
\end{eqnarray}
This T-dual description means that the open string is winding around 
the compactified directions by n times ,whose endpoints are attached to 
same D-brane. This fact leads us to a conclusion that the bulk 
part of the open string freely moves in full D dimensional space, the 
boundary parts are however constrained to p+1 dimensional hyper-surface. 
This solved the problem for the bulk part of the string which can be 
viewed as both closed and open string world sheet. 
\par 
As a final work in this section, we will show how string theory with 
Chan-Paton factors (which has U(N) gauge symmetry) can be realized 
in terms of D-branes.
\par
Let X be a compactified coordinate (here an index of the direction is omitted).
We assume that along the direction there is a constant gauge field of the form  
\begin{eqnarray}
A_{ij}=\frac{1}{2 \pi R} {\rm diag}(\theta_1,\cdots ,\theta_N).   \label{2-6}
\end{eqnarray}
Here the string in which we are interested is an oriented one, whose 
charges at end points transform according to the adjoint representation 
$N \times \bar{N}$ of U(N). The action is given by
\begin{eqnarray}
S=-\frac{1}{4 \pi \alpha '} \int d^2 \sigma \partial_{\alpha}X_{\mu}
\partial^{\alpha} X^{\mu} + \int d \tau A_{\mu}
\dot{X}^{\mu} |_{\sigma =\pi} -\int d \tau A_{\mu}
\dot{X}^{\mu} |_{\sigma =0} 
\end{eqnarray}
where $A_{\mu}$ is the gauge field with $A_i=\frac{\theta _i}{2 \pi R}$ 
coupled to the string endpoints. The assumption that the string is oriented 
is responsible for the difference between the signs in front of the terms 
with gauge field. The conserved current (momentum) 
is 
\begin{eqnarray*}
P_\tau ^{\mu}
=\frac{1}{2 \pi \alpha '}\partial_{\tau} X^{\mu} -\frac{1}{2 \pi R}
(\theta _j -\theta _i).
\end{eqnarray*}
The conserved charge is 
\begin{eqnarray*}
p
= \sqrt{\frac{2}{\alpha '}}\alpha_0 
- \frac{1}{2  R}(\theta _j -\theta _i)
\end{eqnarray*}
Similar procedure to (\ref{2-5}) gives
\[
p=\frac{n}{R}.
\]
The distance between two endpoints of the open string in T-dual theory is 
\begin{eqnarray*}
\hat{X}(\pi)-\hat{X}(0)
= -(2\pi n +\theta_j -\theta_i) \hat R .
\end{eqnarray*}
What we can read into this equation is the open string is attached to 
different D-branes whose coordinates are $\theta_i \hat{R}$ and 
$\theta_j \hat{R}$, respectively. (\ref{2-6}) gives
\[
\theta_i \hat{R} =2 \pi \alpha ' A_{ii},
\]
then i-th D-brane coordinate is written as
\[
X_i=- 2 \pi \alpha ' A_{ii}.
\]
The U(N) gauge field represents the D-brane's coordinates. From this result, 
one can say that a string with Chan-Paton factors is equivalent to a 
string whose boundary points are constrained in D-branes. 
%
%
\section{Dirichlet Condition and Momentum Flow}
It is well known that there is no momentum flowing out of open string's 
end points for the Neumann boundary condition. The reason is the 
following. 
\par
By the N\"other method the current with respect to the translation on the 
worldsheet (i.e. momentum) is given by
\begin{eqnarray*}
P_{\alpha} ^{\mu} =T \partial_{\alpha} X^{\mu},
\end{eqnarray*}
where T is the string tension. The momentum flow across a line 
segment $d \tau$ on the world sheet is given by
\begin{equation}
dP^{\mu}=P^{\mu} _{\sigma} d \tau.   \label{3-1}
\end{equation}
The Neumann boundary condition implies that there is no momentum 
flowing out of the end of the string. 
\par
On the other hand, the open string with the Dirichlet boundary 
condition does not have the same property. This string satisfies 
the condition
\[
\partial_{\tau} X^{\mu} |_{\sigma=0, \pi}=0.
\]
Therefore the momentum flow (\ref{3-1}) does not vanish. This flow into 
the D-brane to which the open string attaches. Hence the 
D-brane has the momentum, in other words, the D-brane is 
dynamical object.
%
%
\section{Bosonic Boundary State}
The interaction between two Dp-branes is described by the vacuum fluctuation of an 
open string which is between them. Its lowest order contribution is the open string 
1-loop diagram which is illustrated by a cylinder. By exchanging the role of the 
world sheet coordinates $\tau$ and $\sigma$ this open string amplitude can also be 
viewed as a closed string tree amplitude which propagates between the Dp-branes. 
Here we would like to make use of this idea. 
\par
Our main purpose in this section is to construct a boundary 
state \cite{callanbeta,callanloop} which describes boundary conditions of a closed 
string which is attached to a Dp-brane. 
At a boundary $\tau =0$ one define the boundary state as 
\begin{eqnarray}
\partial{\tau} X^{\alpha}|_{\tau=0} |B \rangle&=&0 \hspace{1cm} 
 \alpha=0,1, \cdots ,p     \label{4-6} \\
X^{i} |_{\tau=0}|B \rangle &=&y^i|B \rangle  \ \ \ \ \ \ \ \ i= p+1, \cdots ,D-1.
                                    \label{4-7}
\end{eqnarray}
Analogous conditions are hold at another endpoint of the string. 
In order to satisfy the boundary conditions (\ref{4-6}) and (\ref{4-7}) for arbitrary 
$\sigma$, the following expressions have to be hold;
\begin{eqnarray}
(\alpha _n ^{\alpha} +\tilde{\alpha} _{-n} ^{\alpha})|B \rangle=0  \label{4-3}    \\
(\alpha _n ^{i} -\tilde{\alpha} _{-n} ^{i})|B \rangle =0         \label{4-4}        \\
p^{\alpha}|B\rangle=0                           \label{4-1} \\
(x^i-y^i)|B\rangle=0.                      \label{4-2}
\end{eqnarray}
These expressions for the non-zero mode are also written as
\begin{eqnarray*}
(\alpha _n ^{\mu} +S^{\mu} _{\ \nu} 
\tilde{\alpha} _{-n} ^{\nu})|B\rangle=0,
\end{eqnarray*}
here we introduced the matrix 
\[
S^{\mu \nu} = (\eta ^{\alpha \beta} , -\delta ^{ij} ).
\]
\par
So far we have found the conditions which the boundary state for a D$p$-brane must 
satisfy. The above expressions are adequate for investigation of the non-commutativity 
in string theory and we have no need to know the explicit form of the 
boundary state. It is easy problem, however, to write down the boundary state itself. 
The answer is 
\begin{eqnarray}
|B \rangle =N_p \delta^{d-p-1}(x^i -y^i)  \exp(-\frac{1}{n} \alpha_{-n}
\cdot S \cdot \tilde{\alpha} _{-n})  |0 \rangle     \label{4-5}
\end{eqnarray}
where $N_p$ is a normalization factor, and $|0\rangle$ is a ground state 
with respect to the operators $\alpha, \tilde{\alpha},p$.
\par
For the zero-modes it is trivial for the boundary state to satisfy
the conditions (\ref{4-1}) and (\ref{4-2}). Let's check others (\ref{4-3}) and 
(\ref{4-4}). The commutation relation for the oscillators is given by 
\[
[\alpha_m,\alpha_n]=m\delta_{m+n}\eta^{\mu \nu}.
\]
The counterpart for $\tilde{\alpha}$ is the same form.
In the following the index of $\alpha_n$ are $n>0$\footnote{These are
annihilation operators.}. It is easy to see that 
\[
[\alpha_n ^{\mu},e^{-\frac{1}{n}  \alpha_{-n}
\cdot S \cdot \tilde{\alpha} _{-n} } ]=-S^{\mu} _{\ \nu}
\tilde{\alpha}^{\nu} _{-n}
e^{-\frac{1}{n}  \alpha_{-n}
\cdot S \cdot \tilde{\alpha} _{-n} }.
\]
We can read into this equation that the boundary state 
(\ref{4-5}) satisfies boundary conditions (\ref{4-3}) and (\ref{4-4}). Here 
we do not determine the normalization factor$N_p$, but 
this is very important when, for example, one calculate a   
beta function with respect to open string theory with background 
fields.  
\par
In this section we explained the basics of the boundary state in the case that there 
are not background fields. we will see the boundary states 
with constant background fields, when we review the Ishibashi's 
papers \cite{ishibashibrane,ishibashirelation} in the next chapter in which we see 
the equivalence between a Dp-brane and $\infty$ D(p-2)-branes.


\chapter{D$p$-Brane from D$(p-2)$-Branes}
\section{Classical Solution of the Theory of D-Branes}
Lower dimensional D-branes such as D-particles and D-instantons are used to construct 
the matrix models as the fundamental degrees of freedom\cite{banks,ishibashimatrix}. 
N. Ishibashi et.al.\cite{ishibashimatrix} showed that a D-string can be expressed as 
a classical configuration of infinitely many D-instantons in the \greek2B matrix 
model. In the BFSS matrix theory\cite{banks} the conjecture is that M-theory in the 
infinite momentum frame is equivalent to the $N \to \infty $ limit of the theory 
of N D0-branes. P.K. Townsend suggested that a classical supermembrane 
configuration could be identified with D0-branes\cite{townsend}. 
The purpose in this chapter is to generalize this to the relation between D$p$-brane 
and 
D$(p-2)$-branes with $p>1$ in the bosonic string theory. The consideration leads to 
the equivalence between non-commutative
and commutative field theories\cite{ishibashibrane,ishibashirelation}. 
\par
An open string which is attached to a D$p$-brane  has two kinds of massless modes;
a gauge field on the D-brane $A_{\alpha} (\xi)$ $(\alpha =0,1,\cdots ,p)$ 
and  corrective coordinates $M_i (\xi)$ $(i=p+1,\cdots,D-1)$.
$M_i (\xi)$ describes the position of fluctuating D$p$-brane. The 
dynamics of the D-brane is represented by these fields. 
\par
If there are N parallel D-branes which are overlapped, the number of the ways the 
oriented open string  attach to the  
D-branes is $N^2$. This is equivalent to the number of massless modes. 
As a result $A_{\alpha}(\xi)$ and $M_i(\xi)$ are promoted to
 $N \times N$ matrices which is first advocated by Witten\cite{witten}. 
\par
This system  is described by p+1 dimensional U(N) Yang-Mills theory with 
U(N) adjoint scalar fields $M^i$;
\beeqno
S=\int d^{p+ 1}\xi {\rm tr}[-\frac{1}{4g_s l_s ^{p-3}} F_{\alpha \beta}
F^{\alpha \beta}-\frac{1}{2g_s l_s ^{p+1}}D_{\alpha}M_i D^{\alpha}M^i
+\frac{1}{4g_s l_s ^{p+5}}[M_i ,M_j]^2]
\eneqno
where 
\beeqno
D_{\alpha}M_i=\partial_{\alpha}M_i -ig_s[ A_{\alpha},M_i],
\eneqno
$g_s$ is a string coupling constant and $l_s$ is a string length. 
Under the condition that the gauge field vanish and $M^i$ is static, 
the action becomes
\beeqno
S=\frac{1}{4g_s l_s ^{p+5}}\int d^{p+ 1}\xi
{\rm tr}[M_i ,M_j]^2.
\eneqno
This has an equation of motion;
\beeqno
[M_i,[M_i,M_j]]=0.
\eneqno
This has a trivial solution;
\beeqno
[M_i,M_j]=0
\eneqno
which implies that the spacetime is an ordinary commutative one. 
However there is also a non-trivial solution;
\beeq
[M_i,M_j ]= i \theta _{ij},               \label{5-1}
\eneq
where $\theta_{ij}$ is a constant antisymmetric tensor times $N \times N$ unit 
matrix. And this 
expression is valid only for the case that $X^i$ is $\infty \times \infty$ 
matrix. In other words, the open string  
has infinitely many Chan-Paton factors. 
One can understand this fact by operating a trace on the both 
sides of (\ref{5-1}).  Hence the string theory contains automatically 
a non-commutative structure of a spacetime. 
\par
It is easy to naively understand that the theory of $infty$ D$p$-branes with 
the background (\ref{5-1}) may be identical to the theory of D$(p+2)$-brane. 
Transverse directions of D$p$-brane is non-commutative. This causes the uncertainty 
$\delta X ^i \delta X^j \sim \theta ^{ij}$. If one might represent the configuration 
of D5-branes as $(1,2,3,4,5,\times,\times ,\times ,\times)$, we can't detect the 
precise position of the D5-brane along $X^6,\cdots,X^9$ directions by the effect of 
the uncertainty.  Because of this, it can 
be viewed as if the D$p$-brane is extended to these directions. Therefore one 
can consider the branes as a D$(p+2)$-brane effectively. In section 3.3 and 3.4 
we will see this more precisely in the framework of  the boundary state formalism.
\section{Boundary Condition of an Open String}
The coupling term with gauge field in the bosonic string action 
takes the form
\beeq
S_A =\int d \tau A_{\mu} \dot{X}^{\mu}.      \label{5-3}
\eneq
One can change the line integral to the surface integral form using 
Stokes theorem;
\beeqno
\int_{\partial C} \omega =\int _{C} d \omega.
\eneqno
In our case, $\partial C$ is a path on which an  end point of an open string 
is, and $\omega =A_{\mu} dX^{\mu}$. 
Hence $S_A$ is rewritten as 
\beeq
S_A=\frac{1}{2}\int d \tau d \sigma 
F_{\mu \nu}\epsilon ^{ab}
\partial_{a} X^{\mu} \partial_{b} X^{\nu}.   \label{5-16}
\eneq
In this form of the action U(1) gauge invariance is manifest.
We would like to know how the Neumann boundary 
condition is deformed by the effect of the gauge field. 
If $F_{\mu \nu}$ is not constant, 
the equation of motion will be modified from the ordinary free field equation.
The form of string's boundary 
condition doesn't depend on whether $F_{\mu \nu}$ is constant or not. 
The boundary condition is 
\beeqno
\partial_{\sigma} X_{\mu} +2 \pi \alpha ' \partial_{\tau}X^{\nu}F_{\mu \nu} =0
\eneqno
at boundaries. In the closed string picture \footnote{to which is referred as the tree 
channel in \cite{polchinski}} 
the role of $\tau$ and $\sigma$ are exchanged.
\section{The Boundary State}
A boundary state corresponding to a D$(p-2)$-brane which is at $X^i =0$ is 
defined by
\beeqno
\left|B \right>_{p-2} =\left|X=0\right> \otimes \left| B \right>_{gh}
\eneqno 
where the state $\left |X=f\right >$ satisfies
\beeqno
X^i(\sigma)\left |X=f\right >=f^i (\sigma)\left |X=f\right >,
\eneqno
and $\left |B\right >_{gh}$ is a ghost part which is included so that the 
boundary sate may be BRS invariant. This part is  neglected in this section  
since this is not important in our purpose. A boundary state 
which describe a system of N D$(p-2)$-branes with background (\ref{5-1})
is written in terms of a 
Wilson line factor as 
\beeq
\left |B\right >_{N(p-2)}=\tr{\rm  P} \exp(-i \int ^{2 \pi} _0 d \sigma 
P^{i} (\sigma)M_i)\left |B\right >_{p-2} ,         \label{5-2}
\eneq
where P denotes a path ordering with respect to the path $\sigma$. $P^i$ is the 
conjugate momentum with respect to the string coordinates 
($P^i =-\frac{1}{2 \pi \alpha '}\partial_0 X^i$) and $M^i$ satisfies the commutation 
relation (\ref{5-1}). In this chapter, we are interested in the boundary 
state $\left |B\right >_{N(p-2)}$ with $N=\infty$. 
\par
Changing the Wilson line factor to the path integral 
representation is our next task.
In this work it is the key aspect that the coordinates of the D-brane satisfy the 
commutation relation (\ref{5-1}). It is helpful to deal with this system in analogy 
with the quantum mechanics. We deal with the $M^i$ as operators. The eigenvector of 
$M^i$ is denoted by $\left |y^i\right >$;
\beeqno
M^i |y^i\rangle =y^i  |y^i \rangle.
\eneqno
In this equation repeated indices are unsummed. The $y^i$ 
representation of $M^j$ is 
\beeqno
M^j =i \theta ^{ji} \frac{\partial}{\partial y^i}. 
\eneqno
In the representation space (\ref{5-1})
up to normalization $y^i$ representation of the eigenstate of $y^j$ is
\beeqno
\langle y^i  | y^j \rangle \sim \exp [i \omega^{ij}y^i y^j]
\eneqno
where the $\omega$ is inverse of $\theta$.
We will restrict ourselves to the case in which the value of the 
parameter $\theta$ is 
\beeqno
\theta^{kl}&=&\theta  \hsp1 ( k=p-1,l=p) \\
\theta^{ij}&=&0    \hsp1   (i,j= {\rm others}).
\eneqno
\par
Let us divide the path $\sigma$ in the Wilson line factor in (\ref{5-2}) into 
N points with $\sigma_{i+1}-\sigma_i=\epsilon$ ($\epsilon$ is 
infinitesimal constant). The full Wilson line is from $\sigma=0$ to $\sigma= 2 \pi$. 
We denote the Wilson lines which is from $\sigma_i$ to $\sigma_{i+1}$ as
$W(i+1,i)$. The full Wilson line factor $W(2\pi,0)$ is decomposed as
\beeqno
W(2\pi,0)&=&{\rm trP} 
\exp(-i \int ^{2\pi} _0 P_i M^i d \sigma) \\
&=& {\rm tr} W(2 \pi, N)W(N, N-1) \cdots W(1, 0)   
\eneqno
with
\beeqno
W(i+1,i)=\exp [-i \epsilon P_iM^i].
\eneqno
Performing the trace gives  the path integral representation
of the Wilson line factor.
Hence $\infty$ D$(p-2)$-branes boundary state is written as
\beeq
|B \rangle _{N(p-2)} =\int \cald y\exp [i \omega \int d \sigma y^{p-1} 
\partial_{\sigma} y^{p}
-i \int d \sigma (P_{p-1} y^{p-1}+P_{p} y^{p})] |B \rangle _{p-2}.    \label{5-7}
\eneq
Next, we would like to see what conditions the $\infty$ D($p-2$)-branes boundary state 
$\left |B\right >_{\infty(p-2)}$ satisfy. We use the fact an integral of a total 
derivative vanish.
\beeqno
0&=& \int \cald y \frac{\delta}{\delta y^{p-1}}
\exp[i \omega\int d \sigma y^{p-1} 
\partial_{\sigma} y^{p}  -i \int d \sigma (P_{p-1} y^{p-1}+P_{p} y^{p})]
\left |B\right >_{p-2}  \\
&=&[i \omega \partial_{\sigma}X^{p}-iP^{p-1} ]
\left |B\right >_{N(p-2)} 
\eneqno
This implies that the $\infty$ D$(p-2)$-branes boundary state satisfies 
the condition of a D$p$-brane boundary state with gauge 
field $F^{p-1\ p}=\omega$. Hence one can say that the configuration 
of N D$(p-2)$-branes with 
$N= \infty $ is equivalent to that of a D$p$-brane with gauge field. 
\section{Worldvolume Theory}
In the last section, we have explored the case in which the D$p$-branes 
are flat which correspond 
to the D-brane without scalar fields $\phi^i$. Here we would like 
to include effects of the collective coordinates.
\subsection{D$p$-Brane Picture}
In the open string spectrum the collective coordinates of a D-brane
 correspond to scalar fields $\phi ^i$ ($i=p+1,\cdots,D-1$). The 
boundary state with the scalar fields is written as 
\beeqno
\left |B\right >_{p,\phi}=\exp(-i \int ^{2 \pi} _0 d \sigma P_i \phi^i)
\left |B\right >_{p}.
\eneqno
Next we construct a boundary state with a gauge field background 
\cite{okuyama,abouelsaood}. At first, we introduce the coherent state $|x\rangle$ 
which satisfies 
\beeqno
X^i (\sigma) |x \rangle =  x^i(\sigma) |x \rangle
\eneqno
This is described as
\beeqno
|x \rangle = \exp (-i \int d \sigma P_i (\sigma)x^i (\sigma)) |D \rangle 
\eneqno
where the state $|D \rangle$ is defined by the condition 
$X^i (\sigma)|D \rangle=0$.
The boundary state with $U(1)$ gauge field is 
\beeqno
|B \rangle_{pA} &=& \int \cald y \exp(i \int A(y))|y \rangle.  \\
&=& \int \cald y \exp(i \int A(y) 
-i \int d \sigma P_i (\sigma)y^i (\sigma)) |D \rangle .
\eneqno
\par
In the D$p$-brane picture, the open string mode is $A_{\alpha}$ and $\phi^i$. In 
these backgrounds the 
boundary state is guessed to be 
\beeq
\left |B\right >_{A,\phi,p}=\int {\cal D}y 
\exp[i \int d \sigma A_{\alpha}\partial_{\sigma}y^{\alpha}
-i\int d\sigma(P_{p-1} y^{p-1}+P_{p} y^{p}+ P^i \phi_i)]
\left |B\right >_{p-2} \label{5-8}
\eneq
which is coincide with (\ref{5-7}) when $F_{p-1,p} = \omega ,\phi^i=0$.
Small variations $\delta A_{\alpha},\delta\phi^i$ from the backgrounds 
$F_{p-1 \ p}=\omega,\phi^i=0$ in (\ref{5-7}) can be described by acting the following 
vertex operator on $|B \rangle _{N (p-2)}$
\beeq
(1+i \int d \sigma (\delta A_{\alpha} \partial_{\sigma}X^{\alpha}
-\delta \phi^i P_i))|B \rangle _{N(p-2)}      \label{5-9}
\eneq
which is consistent with (\ref{5-8}).
It is easy to notice that because of the part \\
$\exp(-i \int ^{2 \pi}_0 d\sigma (P_{p-1} y^{p-1}+P_{p} y^{p} +P^i \phi_i))$ 
the above boundary state corresponds to the one \\
$\left| X^{p-1}=y^{p-1},X^{p}=y^{p},X^i=\phi^i \right>$. 
This corresponds to the background
\beeq
X^{p-1}=M^{p-1}     \hsp1  X^{p}=M^p   \hsp1
M^i =\phi ^i (X^{\alpha},M^{p-1},M^p).       \label{5-12}     
\eneq
\subsection{D$(p-2)$-Brane Picture}
We can parameterize the background more generally as
\beeqno
X^{\mu}=\phi^{\mu}(X^{\alpha},M^{p-1},M^p),
\eneqno
in which $X^{\alpha},M^{p-1}$ and $M^p$ play a role of the coordinates on the 
worldvolume of D$p$-brane. If we regard the worldvolume as that of D$(p-2)$-brane, 
the worldvolume theory is non-commutative. The boundary state 
with such background is 
\beeq
\left |B\right >_{p \phi}=\int {\cal D}y 
\exp[i \omega \int d \sigma y^{p-1} \partial_{\sigma}y^{p} -i
\int ^{2 \pi}_0 d\sigma P^{\mu} \phi_{\mu}]\left |B\right >_{p-2}.  \label{5-5}
\eneq
\par
In the D$p$-brane picture, on the worldvolume there are $A_{\alpha} \ 
(\alpha =0,\cdots,p)$ and $\phi^i \ (i=p+1,\cdots, D-1)$. On the other hand, 
in the D$(p-2)$-brane picture the worldvolume fields are 
$A_{\alpha} \ (\alpha=0,\cdots,p-2)$  and $\phi^i \ (i=p-1,\cdots, D-1)$. In 
D$(p-2)$-brane picture, there are not $A_{\alpha} \ (\alpha=p-1,p)$. And in the 
D$p$-brane picture, there are not  $\phi^i \ (i=p-1,p )$. Here we would like to 
see the relation between these fields.
Using (\ref{5-9}) variations $\delta A_{\alpha}$ and $ \delta \phi^i$ in 
$|B \rangle _{N(p-2)}$ is written as
\beeq
\delta |B \rangle _{N(p-2)} =i \int \cald y \delta A_{\alpha} 
\partial_{\sigma} y^{\alpha}
\exp [i \omega \int d \sigma y^{p-1} \partial_{\sigma} y^{p}
-i \int d \sigma (P_{p-1} y^{p-1}+P_{p} y^{p})] |B \rangle _{p-2} \nonumber    \\
                                         \label{5-10}
\eneq
and
\beeq
\delta |B \rangle _{N(p-2)} =-i \int \cald y \delta \phi^i P_i
\exp [i \omega \int d \sigma y^{p-1} \partial_{\sigma} y^{p}
-i \int d \sigma (P_{p-1} y^{p-1}+P_{p} y^{p})] |B \rangle _{p-2}   \nonumber  \\   
                                                       \label{5-11}
\eneq
respectively. Using the identity
\beeqno
0&=&\int {\cal D}y \frac{\delta}{\delta y ^{\alpha}}\exp[i
\int d\sigma A_{\beta}\partial_{\sigma}y^{\beta}-i
\int d \sigma(P_{p-1} y^{p-1}+P_{p} y^{p}+ P_{i}\phi^{i}]
\left |B\right >_{p-2}    \hsp1 \\
&=&\int {\cal D}y [iF_{\alpha \beta}\partial_{\sigma}y^{\beta}
-iP_{\alpha}]\exp[i
\int d\sigma A_{\alpha}\partial_{\sigma}y^{\alpha}-i
\int d \sigma (P_{p-1} y^{p-1}+P_{p} y^{p}+P_{i}\phi^i)]\left |B\right >_{p-2} 
\eneqno
(\ref{5-10}) and (\ref{5-11}) are coincide each other if 
$\delta A= \omega \delta \phi$.
When the static gauge is used, the relation becomes 
\beeq
\delta A= \omega \delta y   \label{5-15}
\eneq
which implies that the reparametrization $\delta y$ is equivalent to
the variation of the gauge field.  
\par
The relation between D$p$-brane picture and D$(p-2)$-brane picture can be viewed from 
another point of view. We consider the boundary state involving all fields 
$A^{\alpha}$ and $\phi^i$. Since there are extra fields there must be symmetry to 
reduce the extra degrees of freedom. This is reparametrization invariance. 
If we use the static gauge, the boundary state becomes that of D$p$-brane picture. 
On the other hand the gauge condition $F_{p-1 \ p}=\omega$ correspond to 
the D$(p-2)$-brane 
picture. 
\par  
We summarize the result. We observed that a D$p$-brane with gauge field $F=\omega$ 
can be viewed as a configuration of infinitely many D$(p-2)$-branes. 
If we consider the system as 
one D$p$-brane, the worldvolume theory is the  gauge theory on ordinary commutative 
space. 
However in the D$(p-2)$-brane picture the worldvolume theory is non-commutative gauge 
theory. 
\par
Let us see about the symmetry. In the D$(p-2)$-brane picture the deformation of a 
configuration of D-branes along $X^{p-1}$ and $X^p$ directions is parameterized by the  
scalar fields $\phi^i$. The space of deformation would be 
\beeq
\frac{{\rm space \ of}\  \phi^i}{{\rm Diff}_F}              \label{5-13}
\eneq 
where $Diff_F$ is the group of diffeomorphism which preserve the field 
strength. 
The deformation of a D$p$-brane along $ X^{p-1}$ and $X^p$ directions 
is also parameterized by $A_{\alpha}$ . The counterpart of 
(\ref{5-13}) is
\beeq
\left( \frac{{\rm space \ of \ A_{\alpha}}}{G} \right)  \label{5-14}
\eneq
where G is the gauge group. (\ref{5-13}) and (\ref{5-14}) are equivalent to 
each other by virtue of the relation $\delta A=\omega \delta \phi$.
\section{Extension to multiple D$p$-branes}
In the last section we showed that a D$p$-brane with a constant gauge field strength 
is equivalent to a configuration of 
infinitely many D$(p-2)$-branes. Here we extend this result to the relation 
between n D$p$-branes (n is finite) and $\infty$ D$(p-2)$-branes. On the 
worldvolume of n D$p$-branes the gauge field and the scalar field are in 
adjoint representation of U(n) group. In the D$p$-brane picture the 
boundary state is written as
 \beeq
\left |B\right >_{A,\phi,p}=\int {\cal D}y \trp 
\exp[i \int d \sigma A_{\alpha}\partial_{\sigma}y^{\alpha}
-i\int d\sigma(P_{p-1} y^{p-1}+P_{p} y^{p}+ P^i \phi_i)]
\left |B\right >_{p-2}                     \label{9-1}
\eneq
where $A_{\alpha}$ and $\phi^i$ are $n \times n$ matrices. The difference 
between (\ref{9-1}) and (\ref{5-8}) is that there is $\trp$ in 
(\ref{9-1}). As we did in section 3.3, we would like to make the factor
$\trp$ easy to deal with.  The technique to deal with  $\trp$ is 
suggested by 
Samuel\cite{samuelfunctional}. The technique is applied by many 
authors\cite{halpern,samuelcolor,brandt,arefeva,gervais,dorndual,dornnonabelian,dornopen}. 
\par
Let $H=iA_{\alpha}\partial_{\sigma}y^{\alpha}
-i(P_{p-1} y^{p-1}+P_{p} y^{p}+ P^i \phi_i)$
and $\sigma_{j+1}-\sigma_{j}=\epsilon$ $(j=0,1,\cdots, 2 \pi)$.
We introduce a set of anti-commuting variables 
$\eta_{l}(\sigma_j)=\eta_{l}(j)$ and 
$\eta^* _{l}(\sigma_j)=\eta^* _{l}(j) \  
(l=1, \cdots, n)$ which satisfies anti-commutators 
\beeqno
[\eta_l(i),\eta_k(j)]^{(+)}=0,     \hsp1
[\eta_l^{*}(i),\eta_k(j)]^{(+)}=0 {\rm \ and}    \hsp1
[\eta^* _l(i),\eta^* _k(j)]^{(+)}=0.   
\eneqno
Consider 
\beeq
T_{lk} \equiv \int d\eta(2 \pi)d\eta^*(2\pi)\int d\eta(N)d\eta^*(N) \cdots 
\int d\eta(0)d\eta^*(0) \eta_l ( 2\pi) \exp(\sum_{j=o}^N C_j)\eta^* _k(0) 
                                           \label{9-2}
\eneq
where $\int d\eta(i)d\eta^*(i)=\prod_k \int d\eta_k(i)d\eta^* _k(i)$ and 
\beeq
C_j \equiv C(\sigma_j) \equiv \sum_k [\eta_k^* (j+1)-\eta^*_k (j)]\eta_k (j)
+\sum_{il}\eta^* _i(j+1)H_{il}(x_j)\eta_l(j) \epsilon
\eneq
We would like to show 
\beeqno
\trp \exp[\int d \sigma H]=\tr T_{lk}
\eneqno
In the exponential in (\ref{9-2}) there is a factor; 
\beeq
\exp[\sum_k \eta_k (j)\eta^* _k (j)]=1+\sum_k \eta_k (j) \eta ^* _k (j)        
+\cdots +\frac{1}{n !}(\sum_k \eta_k (j) \eta ^* _k (j))^n.   \label{9-3}
\eneq     
Integrations of (\ref{9-3}) multiplied with 
\begin{eqnarray}
   \left\{ 
           \begin{array}{c}
                             1        \\
                          \eta_k (j) \eta ^* _k (j)   \\
\eta_{k_1} (j) \eta ^* _{k_1} (j)\eta_{k_2 }(j) \eta ^* _{k_2} (j) \hsp1
  (k_1 \neq k_2) \\
              \vdots     \\
\eta_{1} (j) \eta ^* _{1}(j)\cdots   \eta_{k-1 }(j) \eta ^* _{k-1} (j)
            \eta_{k+1} (j) \eta ^* _{k+1} (j)  \cdots 
            \eta_{n} (j) \eta ^* _{n} (j)   \\
       \eta_{1} (j) \eta ^* _{1} (j) \cdots \eta_{n}(j) \eta ^* _{n} (j) 
                    \end{array} 
                         \right.     \label{9-4}
\end{eqnarray} 
respectively with respect to $\eta(i),\eta^*(i)$ are equivalent to one. 
All other integrations are zero. Another exponential factor is  
\beeqno
\exp \{\sum_{k} \eta_k^* (j+1)\eta_k (j)
+\sum_{il}\eta^* _i(j+1)H_{il}(x_j)\eta_l(j) \epsilon  \}.
\eneqno
Expand this and perform $\eta (0),\eta^* (0)$ integrals in (\ref{9-2});
\beeqno
&& \int d\eta(0)d\eta^*(0)\exp(\sum \eta(0) \eta^* (0) )\sum_n \frac{1}{n!}
[\sum( \eta_k^* (1)\eta_k (0)
+\eta^* _i(1)H_{il}(x_j)\eta_l(0) \epsilon) ]^n\eta^* _l(0)  \\
&=& \int d\eta(0)d\eta^*(0)\exp(\sum \eta(0) \eta^* (0) )
[\sum( \eta_k^* (1)\eta_l (0)
+\eta^* _i(1)H_{il}(x_j)\eta_l(0) \epsilon) ]\eta^* _k(0)  \\
&=&\sum_k \eta^* _k (1)[I+\epsilon H(1)]_{kl}
\eneqno
where we have used (\ref{9-4}) and $I$ is an unit matrix. Repetition of 
these integrations in (\ref{9-2}) gives 
\beeqno
T_{ij}=\{ [I+\epsilon H(N)]\cdots [I+\epsilon H(1)] \}_{ij}
\eneqno
Using this result the boundary state (\ref{9-1}) is rewritten as
\beeq
&&\left |B\right >_{A,\phi,p}=\int {\cal D}y \cald \eta \cald \eta^*  \nonumber \\
&&\times\exp\{\int d \sigma \eta^* \partial_{\sigma}\eta
+ i\int d \sigma \eta^{*}[A_{\alpha}\partial_{\sigma}y^{\alpha}
-i\int d\sigma(P_{p-1} y^{p-1}+P_{p} y^{p}+ P^i \phi_i)]\eta \}
\left |B\right >_{p-2}.    \nonumber  \\  
       \label{9-5}
\eneq
In short performing  $\eta,\eta^*$ integrals in (\ref{9-5}) gives 
(\ref{9-1}). 

\section{Diff Invariance and Non-Commutative Gauge Symmetry}
In this section we analyze a flat D$p$-brane with a constant NS B field in the 
Type \greek2 theory and show that the symmetry of the 
worldvolume coordinate can lead to a non-commutative 
gauge symmetry on the D-brane\cite{cornalba}.
The D$p$-brane is extended in the spacetime directions
$X^{\alpha}$ $(\alpha=0 ,\cdots , p)$. The worldvolume coordinates are  
parameterized by $x^{\alpha}$, which are related to the spacetime coordinates as
\beeqno
X^{\alpha}=x^{\alpha}.
\eneqno
We assume that the NS B field has maximal rank $r =p+1$.
For simplicity we restrict ourselves to the type \greek2B theory. 
In this case the space dimensionality $p$ of a 
D$p$-brane is odd and the matrix $B_{\alpha \beta}$ is 
invertible. We denote the inverse matrix of B as $\theta$.
The effective theory of D-brane in the approximation 
that the derivative of the field strength $f_{\alpha \beta}$ of U(1) gauge field is negligible 
is the Dirac-Born-Infeld action. The DBI action has two types of gauge symmetry. 
One of which is ordinary one;
\beeq
A_{\alpha} \to A_{\alpha}+\partial_{\alpha} \lambda.    \label{8-5}
\eneq
And another is 
\beeqno
B_{\alpha \beta} &\to& B_{\alpha \beta} +\partial_{\alpha} \Lambda_{\beta}-
\partial_{\alpha} \Lambda_{\beta}     \\
A_{\alpha} &\to& A_{\alpha}+\Lambda_{\alpha}.
\eneqno
In the action the gauge invariant 
combination of the U(1) gauge field and NS B field is 
\beeqno
F_{\alpha \beta}(x) \equiv B_{\alpha \beta}+f_{\alpha \beta}(x)
\eneqno
where
\beeqno
f_{\alpha \beta}=\partial_{\alpha}A_{\beta}-\partial_{\beta}A_{\alpha}.
\eneqno
\par
In the DBI action, we can fix the $\Lambda$ gauge symmetry as 
$F_{\alpha \beta} =B_{\alpha \beta}$.We transform the worldvolume coordinate $x$ to 
$\sigma$. This transformation causes a change of the total field strength;
\beeqno
\tilde{F}_{\alpha \beta}(\sigma)=\frac{\partial x^{\delta}}{\partial \sigma^{\alpha}}
\frac{\partial x^{\gamma}}{\partial \sigma ^{\beta}} F _{\delta \gamma}(x) .
\eneqno
Let us consider an infinitesimal transformation;
\beeq
x^{\alpha}=\sigma ^{\alpha} +d^{\alpha}.           \label{b}
\eneq
The total field strength becomes
\beeqno
\tilde{F}_{\alpha \beta}
&=& B_{\alpha \beta}+B_{\rho} \partial _{\alpha} d^{\rho} 
+B_{\alpha \rho } \partial _{\beta} d^{\rho}
\eneqno
The coordinate system $\sigma^{\alpha}$ in which the total field strength is 
equivalent to the NS B field is not unique. Coordinate translation from the 
$\sigma^{\alpha}$ to $\sigma ^{\alpha} + V^{\alpha} (\sigma)$, $V^{\alpha}$ 
is infinitesimal, 
with $V^{\alpha}$ satisfying 
\beeqno
B_{\delta \beta} \partial_{\alpha} V^{\delta} +B_{\alpha \delta} \partial_{\beta} V^{\delta}=0,
\eneqno
the NS B field (in other words, total field strength) is invariant. 
$V^{\alpha}=\theta ^{\alpha \beta} \partial _{\beta} \rho$ for arbitrary scalar $\rho$ satisfies 
this equation.
The consequence in this paragraph is that the worldvolume 
with the total field strength $\tilde{F} = B$ has symmetry;
\beeq
x^{\alpha} 
\to  x^{\alpha} +i \{ \rho , x^{\alpha} \}.        \label{8-3}
\eneq
Here we deal with the worldvolume of the D-brane as a symplectic 
manifold with a Poisson bracket
\beeqno
\{ A , B \} \equiv i\theta^{\alpha \beta} \partial_{\alpha} A \partial_{\beta} B
\eneqno
where the A and B are functions depending on canonical variables. 
Using the definition of the Poisson bracket gives 
\beeqno
\{\sigma ^{\alpha} , \sigma ^{\beta} \} = i \theta ^{\alpha \beta}.
\eneqno
We see the meaning of this equation. 
Define a field 
\beeqno
\hat{A}_{\alpha}= B_{\alpha \beta} d^{\beta} (\sigma).
\eneqno
In terms of the gauge field $A$, the symmetry (\ref{8-3}) is written as
\beeq
\hata_{\alpha} &\to& \hata_{\alpha} +\partial_{\alpha} \rho 
+i \{ \rho, \hata_{\alpha} \}.  \label{8-4} 
\eneq
where (\ref{b}) and (\ref{8-3}) were used.
Let us see a relation between the Poisson bracket and a canonical 
commutator on a non-commutative space. If the inverse of the NS B 
field 
$\theta$ is small, one can expand a Moyal product 
factor. From this consideration, it is found that the commutator 
is equivalent to the Poisson bracket up to first order in $\theta$. 
\beeqno
[A,B]_{\theta}=\{A,B \} + {\cal O}(\theta^2)
\eneqno
Hence the symmetry (\ref{8-4}) can be interpreted as the gauge 
symmetry on the non-commutative space which has non-commutative parameter 
$\theta$;
\beeqno
[\sigma^{\alpha} ,\sigma ^{\beta}]_{\theta} =i \theta ^{\alpha \beta},
\eneqno
explicitly
\beeqno
\hata_{\alpha} &\to& \hata_{\alpha} +\partial_{\alpha} \rho 
+i [ \rho, \hata_{\alpha} ]_{\theta}.
\eneqno
The result is essentially equivalent to the result (\ref{5-15}) in section 3.4.   


\chapter{Moyal Product and Non-Commutative Field Theory}
\section{Non-Commutative Geometry}
In string theory the space-time non-commutativity is related to the NS B field which  
is investigated by many authors. In this section we would like to show that a product 
of fields on a non-commutative space is represented by a Moyal product 
\cite{connes,aoki,li}. 
We are not interested in a non-commutative space itself, but functions like 
fields and geometrical objects on a non-commutative space.
\par
The concepts of line, surface and space in geometry can be understood 
in terms of $C^*$ algebra. The definition of the $C^*$ algebra is as follows\cite{connesgeo}. \\
\\
{\it * algebra}  \\
\\
A complex space with multiplication which is associative and 
distributive, and with an involution $a \to a^*$ is called * algebra.
An involution in an algebra ${\cal B}$ is a mapping $a \to a^*$ such that
\beeqno
(T+S)^*  &=&T^*+S^*  \\
(\alpha T)^*&=&\bar{\alpha} T^*  \\
(ST)^* &=&T^* S^*  \\
T^{**} &=&T \\
|| T^* ||&=& ||T||
\eneqno
with $T,S \in {\cal B}$, $\alpha$ is a complex number. 
If $ST=TS$ this is called as commutative * algebra. \\
\\
{\it $C^*$ algebra} \\
\\
$C^*$ algebra is a * algebra with a norm which satisfies the condition 
\beeqno
\|ab\| \leq \|a \| \|b \| , \hsp1 \|a^* a \| = \|a \| ^2.
\eneqno
{\it spectrum}\\
\\
A set SpA of homomorphism $\chi$ of A into $\mathbf {C} $ such that 
$\chi(1)=1$ is compact; the compact space SpA is called the spectrum of A.
There is an important theorem with respect to the {\it commutative} $C^*$ 
algebra. 

\newtheorem{gelfand}{Theorem}

\begin{gelfand}
Let A be a commutative $C^*$ algebra with unit and let $X=Sp A$ be its spectrum.
The Gelfand transformation
\beeq
x \in A \to \chi  \in X             \label{6-3}
\eneq
is an isomorphism of A onto the $C^*$ algebra $C(X)$ of continuous complex function on
X
\end{gelfand}
This implies that a commutative $C^*$ algebra can be described in terms of 
functions. In the next section we will see it's non-commutative version 
and specific identification of a non-commutative $C^*$ algebra with functions 
on a non-commutative space.

\section{Moyal Product}
\par
An operator $\hat{\cal{O}}$ which is an element of a  non-commutative $C^*$ algebra 
can be mapped to an element of a 
function algebra. The Fourier decomposition of the operator $\hat{\cal{O}}$ 
is 
\[
\hat{\cal O}= \int \frac{d^D k}{(2 \pi)^D} e^{i k_{\mu}\hat{x}^{\mu}}\tilde{f}(k)
\]
where the spacetime coordinates $\hat{x}^{\mu}$ satisfy the commutator
\[
[\hat{x}^{\mu},\hat{x}^{\nu}]=i \theta^{\mu \nu}. 
\]
$\theta^{\mu \nu}$ is a constant real anti-symmetric tensor and characterize this 
non-commutative space. A counterpart of (\ref{6-3}) in non-commutative $C^*$ 
algebra is represented by  a Fourier transformation of $\tilde{f}(k)$;
\beeqno
f(x)=\int \frac{dk}{2 \pi}e^{-ikx}\tilde{f}(k).
\eneqno
Fourier transformation of 
$\tilde{f}(k)$ gives a function $f(x)$ on the non-commutative space.
Summarize the process of this.
\begin{eqnarray*}
&\hat{{\cal O}}& \hspace{1cm} : {\rm an \ element \ of \  C^* \ algebra} \\
&\downarrow& \\
&\tilde{f}(k)& \\
&\downarrow&   \hspace{1cm} :{\rm Fourier \ transformation } \\
&f(x)&     \hspace{1cm} : {\rm a \ function \ on  \ the \ non 
\hspace{-.3em}-\hspace{-.3em}commutative \ space } 
\end{eqnarray*}
Secondary let us see how the product of two operators transform.  
\beeqno
\hat{{\cal O}}_f \hat{{\cal O}}_g  \rightarrow \tilde{f}\tilde{g}(k) 
\rightarrow (f*g)(x)
\eneqno
\begin{eqnarray*}
\hat{{\cal O}}_f \hat{{\cal O}}_g 
=\int \frac{d^D k'}{(2 \pi)^D} e^{i k' _{\mu}\hat{x}^{\mu}}\tilde{f}(k')
\int \frac{d^D k''}{(2 \pi)^D} e^{i k'' _{\mu} \hat{x}^{\mu}}\tilde{g}(k'')
\end{eqnarray*}
The product of two operators transformed as
\beeqno
(f*g)(x)=&=&\int \frac{d^D k}{(2 \pi)^D}e^{ikx}
 \int \frac{d^D k'}{(2 \pi)^D}
 e^{-\frac{1}{2}\theta^{\mu \nu}k'  _{\mu}(k -k') _{\nu}}
\tilde{f}(k') \tilde{g}(k-k')      \\
&\equiv&\* f(x+\alpha)g(x+\beta)|_{\alpha=\beta=0}
\eneqno
This product is called ``Moyal product'' or ``star product''.
When $\theta =0$, this is reduced to an ordinary product.
\par
As a consistency check, let us try the case
\beeqno
f(x)=x^1 \hspace{1cm} g(x)=x^2
\eneqno
on a 2 dimensional space ($\theta^{12} \equiv \theta$).
One can readily calculate the commutator of these coordinates whose 
product is the Moyal product.
\beeqno
(f*g)(x)&=&(1+i \theta \frac{\partial}{\partial x^1}
\frac{\partial}{\partial x^2})x^1 x^2   \\
&=&x^1 x^2 +i \theta
\eneqno
This gives
\beeqno
[x^1,x^2]_*=i \theta^{12}.
\eneqno
Hence the map preserves the non-commutativity of the space. 
One of the origins of difficulties of the non-commutative field theory is 
the Moyal product which makes a theory non-local.  
\par
For the Moyal product there are  simple formulae. The first one is
\beeq
\int dx f*g=\int dx g*f.         \label{6-1}
\eneq
We can easily check this equation by seeing order by order in $\theta$. \\
\\
{\it $\theta^0$ order}  
\beeqno
\int fg =\int gf
\eneqno
(Notice that the product is an ordinary one in the equation.)  \\
\\
{\it $\theta^1$ order}  
\beeqno
\int \frac{i}{2}\theta^{\mu \nu}\partial_{\mu}f\partial_{\nu}g
&=&-\int \frac{i}{2}\theta^{\mu \nu}f\partial_{\mu}\partial_{\nu}g \\
&=&0
\eneqno
where in the first line the integration by parts is performed. 
Hence
\beeqno
\int dx f*g=\int dx fg,
\eneqno
and (\ref{6-1}) is satisfied. And the second one is 
\beeq
\delta \theta^{ij}\frac{\partial}{\partial \theta ^{ij}}(f*g)
=\frac{i}{2}\delta \theta^{ij}\frac{\partial f}{\partial x^i}*
\frac{\partial g}{\partial x^j}.   \label{6-5}
\eneq
\section{Non-Commutative Field Theory}
\par
Here we will show some actions of field theory on a non-commutative 
space. we are aware that the only difference with an ordinary theory (on commutative 
space ) is products of fields are the Moyal products which is the result of the 
last section.\\
\\
{\it $\phi ^3$ theory}  \\
\\
The ordinary $\phi^3$ theory is given by
\beeqno
S=\int d^D x \{\frac{1}{2}\partial_{\mu}\phi (x)\partial^{\mu}\phi (x)
+\frac{\lambda}{3!}\phi^3 (x) \}.
\eneqno
Its counterpart in non-commutative theory is 
\beeqno
S&=&\int d^D x \{\frac{1}{2}\partial_{\mu}\phi (x)*\partial^{\mu}\phi (x)
+\frac{\lambda}{3!}\phi*\phi*\phi \} \\
&\equiv&\int d^D x \{\frac{1}{2}\partial_{\mu}\phi (x)\partial^{\mu}\phi (x)
+\frac{\lambda}{3!}\phi^3 (x) \}_*.
\eneqno
\\
{\it $U(1)$ gauge theory}  \\
\\
The ordinary U(1) gauge theory is described by 
\beeqno
S=-\int d^D x F^{\mu \nu}F_{\mu \nu}
\eneqno
where 
\beeqno
F_{\mu \nu}=\partial_{\mu}A_{\nu}-\partial_{\nu}A_{\mu}.
\eneqno
This is invariant under
\beeqno
\delta A_{\mu}=\partial _{\mu}\lambda.
\eneqno
On a non-commutative space, the field strength is 
\beeq
\hat{F}_{\mu \nu}=\partial_{\mu}\hat{A}_{\nu}-\partial_{\nu}
\hat{A}_{\mu}-i [\hat{A}_{\mu},\hat{A}_{\nu}]_*.  \label{6-2}
\eneq
The action is 
\beeqno
S=-\int d^D x \hat{F}^{\mu \nu}*\hat{F}_{\mu \nu}.
\eneqno
The field strength is defined by a commutator of covariant derivatives. 
The last term in (\ref{6-2}) comes from the fact that the gauge fields are 
non-commutative. This theory is U(1) gauge theory, but non-Abelian. 
Let us show that the action is invariant under the transformation
\beeq
\hat{\delta} \hat{A}_{\mu}=\partial_{\mu}\hat{\lambda}
+i [\hat{\lambda},\hat{A}_{\mu}]_*.     \label{6-4}
\eneq
The gauge transformation changes the field strength (\ref{6-2}) as
\beeqno
\delta\hat{F}_{ij}
=i\hatlam * \hatf _{ij} -i\hatf_{ij}* \hatlam.
\eneqno
Under the transformation, although the Lagrangian is not invariant:
\beeqno
\delta (\hatf_{ij}*\hatf^{ij})
= i\hatlam *\hatf_{ij}*\hatf^{ij}-i\hatf_{ij}*\hatf^{ij}*\hatlam,
\eneqno
the action is invariant by virtue of (\ref{6-1}).
\par
In subsequent sections we will show that the low energy effective theory of the 
string with NS B field has non-commutative field theoretical representations.
\section{Conformal Field Theory}
The string theory has a conformal invariance on a string worldsheet. In this 
section
our purpose is to see the symmetry and the operator product 
expansion. We use an Euclidean signature with respect to the string 
worldsheet. Defining complex variables;
\beeqno
\rho &=&\tau+i\sigma   \\
\bar{\rho} &=&\tau-i\sigma,  
\eneqno
the conformal symmetry is defined in terms of $\rho$ and $\bar{\rho}$ as
\beeqno
\rho &\to& f(\rho)   \\
\bar{\rho} &\to& \bar{f}(\bar{\rho})
\eneqno
where $f(\rho)$ and $\bar{f}(\bar{\rho})$ are holomorphic functions. 
The left and right movers are not mixed under the transformation.  
\par 
Define z plane as $z=e^{\rho}$ which is mainly used below. Define a  
correlation function as 
\beeq
\langle X(z)Y(w)\cdots \rangle \equiv \langle \Omega| T[X(z)Y(w)\cdots ]
|\Omega \rangle                  \label{10-2}
\eneq
where $X$ and $Y$ are operators on the z plane $|\Omega \rangle$ is a conformal 
(SL(2,C)) 
vacuum whose definition is given here. The conformal vacuum differ from an 
oscillator 
vacuum which is usually used. A field $\phi(z)$ with a conformal dimension d 
transforms with respect to a coordinate transformation $z\to z'$ as 
\beeqno
\phi'(z')=\left(\frac{dz}{dz'} \right)^d \phi(z).
\eneqno
On the $\rho$ plane oscillator modes $\phi_n$ are defined by 
\beeqno
\phi(\rho) =\sum^{\infty}_{n=-\infty} \phi_n e^{-n\rho}.
\eneqno
On the $z$ plane using $\frac{d \rho}{d z}=\frac{1}{z}$ gives 
\beeqno
 \phi(z)&=&\left(\frac{d \rho}{dz} \right)^d \phi(\rho)    \\
&=&\sum_n \phi_n z^{-n -d}.
\eneqno
$\phi_n$ is given by 
\beeqno
\phi_n = \oint  \frac{d z}{2 \pi i} z^{n+d-1}\phi(z).
\eneqno
Here we demand that $\phi(z)$ is regular at $z=0,\infty$. We consider the case that 
one of the operators in (\ref{10-2}) is $\phi_n$; 
\beeqno
\langle X(z_1)Y(z_2)\cdots \oint _C \frac{dz}{2 \pi \alpha'}z^{n+d-1}\phi(z)\rangle 
\equiv \langle \Omega| T[X(z_1)Y(z_2)\cdots \phi_n]|\Omega \rangle 
\eneqno
We use a small contour C so that $z_1,\cdots$ may not be in the contour. Because 
of the time (radial) ordering, $\phi_n$ is at the right end . Since the integral 
on the left hand side vanishes for $n\geq 1-d$, we have 
conditions for the conformal ket vacuum;
\beeq
\phi_n |\Omega \rangle =0   \hsp1  {\rm for} \ n \geq 1-d.     \label{10-5}
\eneq
Similarly we obtain  conditions for the bra vacuum. 
We have 
\beeq
\langle \Omega | \phi_n =0 \hsp1 {\rm for} \   n \leq d-1 .    \label{10-4}
\eneq
(\ref{10-5}) and (\ref{10-4}) are   definitions of  the conformal vacuum. 
$\phi_n$  with $n \geq 1-d$ are interpreted as annihilation operators, and others 
are creation operators. 
\par
In a conformal field theory operators are usually normal ordered without any 
mention. The 
normal ordering of operators is defined so that modes with $n \geq 1-d$ is on 
the right of 
modes with $\ n \leq 1-d$. For example the normal ordered product $:\phi_n \phi_l :$ with 
$ n \geq 1-d$ and $l \leq 1-d$ is
\beeqno
:\phi_n \phi_l :=\phi_l \phi_n.
\eneqno
Singularity of products of operators  are picked up by using of the operator product 
expansion (OPE) which can be generated from
\beeqno
:X::Y: =:XY:+\sum (\rm cross-contractions)     
\eneqno
for arbitrary operators $X$ and $Y$. Then if we know a propagator, we
can perform the calculation. A practical calculation is given in the next section.   
\section{Open String and Constant NS B Field}
Coordinates of an open string with a background NS B field become 
non-commutative\cite{abouelsaood}.
In this section we examine this in the framework of CFT. 
\par
The string theory has a Diff invariance and the $\Lambda$ gauge invariance as 
discussed above. Here we use the conformal and $F^{\alpha \beta}=0$ gauge fixing 
conditions. The string action with these gauge conditions is 
\beeqno
S=-\frac{1}{4 \pi \alpha'}\int d \tau d \sigma 
(g_{\alpha \beta} \partial_a X^{\alpha} 
\partial^aX^{\beta}-2 \pi \alpha'i B_{\alpha \beta}\epsilon^{ab}\partial_aX^{\alpha}
\partial_bX^{\beta})
\eneqno
where $g_{\alpha \beta}$ and $B_{\alpha \beta}$ are a spacetime metric and a NS B 
field which are constants. Boundary conditions are
\beeqno
g_{\alpha \beta} \partial_{\sigma} X^{\beta} 
-2 \pi \alpha'i B_{\alpha \beta}\partial_{\tau}X^{\beta}=0
\eneqno
at boundaries. Here we consider the case that the string worldsheet $\Sigma$ is 
a disc (in the open string picture this is a tree diagram). The worldsheet 
can be mapped to the upper half plane of the z plane on which the open string  
boundary condition is 
\beeqno
\left.g_{\alpha \beta} (\partial -\bar{\partial}) X^{\beta} 
+2 \pi \alpha' B_{\alpha \beta}
(\partial+\bar{\partial})X^{\beta} \right|_{z=\bar{z}}=0.
\eneqno
The boundaries of the worldsheet are identified with the real axis ($z=\bar{z}$). 
We would like 
to obtain a Green function $G(z,z')$ which satisfies 
\beeq
\frac{1}{2 \pi \alpha'} \partial\bar{ \partial }G(z,z')=-\delta(z-z').   \label{10-1}
\eneq
We use the method of images.
The green function $\langle X^{\alpha}(z)X^{\beta}(z') \rangle$ satisfies 
the boundary condition;
\beeqno
(g_{\alpha \beta}+2 \pi \alpha'B_{\alpha \beta})\partial 
\langle X^{\beta}(z)X^{\gamma}(z') \rangle
=(g_{\alpha \beta}-2 \pi \alpha'B_{\alpha \beta})\bar{\partial} 
\langle X^{\beta}(z)X^{\gamma}(z') \rangle
\eneqno
at $z=\bar{z}$. 
This is 
\beeq
\langle X^{\alpha}(z)X^{\beta}(z') \rangle
&=&-\alpha'
[g^{\alpha \beta} \ln|z-z'|+ \frac{1}{2}\left(
\frac{g-2\pi \alpha' B}{g+2 \pi \alpha'B} \right)^{\alpha \beta}
 \ln(z-\bar{z}')    \nonumber             \\
&&\hsp1 +\frac{1}{2}\left(
\frac{g+2\pi \alpha' B}{g-2 \pi \alpha'B} \right)^{\alpha \beta} \ln(\bar{z}-z')].
                                \label{10-8}
\eneq
We rewrite this in terms of $G$ and $\theta$ which are defined by 
\beeqno
G^{\alpha \beta}&=& \left( \frac{1}{g+2 \pi \alpha'B}g\frac{1}{g-2 \pi \alpha'B}
\right)^{\alpha \beta}    \\
\theta^{\alpha \beta}&=&- 2 \pi \alpha '
\left( \frac{1}{g+2 \pi \alpha'B}B\frac{1}{g-2 \pi \alpha'B}
\right)^{\alpha \beta}.
\eneqno
What we would like to do here is to divide these terms into two parts. One of which 
is single-valued, and others are not. 
Therefore the Green function is 
\beeqno
\langle X^{\alpha}(z)X^{\beta}(z') \rangle
&=&-\alpha[g^{\alpha \beta} \ln|z-z'|-g^{\alpha \beta} \ln|z-\bar{z}'|
+G^{\alpha \beta}\ln |z-\bar{z}'|^2   \\
&&\hsp1 +\frac{\theta^{\alpha \beta}}{2 \pi \alpha '}
\ln \frac{z-\bar{z}'}{\bar{z}-z'}
+D^{\alpha \beta}]
\eneqno
where $D^{\alpha \beta}$ is a  constant. In this chapter we only need 
the Green function at the boundary (${\rm Im} Z=0$). The green function except for the 
fourth term is single-valued. 
We define a sign function as
\beeqno
\epsilon(\tau) \equiv 2 \theta(\tau)-1.
\eneqno
Let $D^{\alpha \beta}$ have a convenient value, at boundary 
$\langle X^{\alpha}(z)X^{\beta}(z') \rangle$ is 
\beeq
\langle X^{\alpha}(\tau)X^{\beta}(\tau') \rangle
=-\alpha' G^{\alpha \beta}\ln (\tau-\tau')^2 
+\frac{i}{2}\theta^{\alpha \beta}\epsilon(\tau-\tau') .   \label{10-6}
\eneq
It is easy to see that $\theta^{\alpha \beta}$ is interpreted as 
a non-commutative parameter. Set $\tau^{\pm}=\tau \pm \epsilon$ 
($\epsilon>0$), the commutator of coordinates of the open string is 
written by using the time ordering operator T;
\beeqno
[X^{\alpha}(\tau),X^{\beta}(\tau)]=T(X^{\alpha}(\tau)X^{\beta}(\tau^-)
-X^{\alpha}(\tau)X^{\beta}(\tau^+)).
\eneqno
Using the Wick theorem gives
\beeqno
[X^{\alpha}(\tau),X^{\beta}(\tau)]
=i \theta^{\alpha \beta}
\eneqno
which implies that the coordinates on the D-brane is non-commutative. 
If there is no background NS B field, $\theta=0$ which means that the 
spacetime is commutative. As we have seen in section 4.2, on non-commutative 
space, a product of functions is the star product.  
Let us see a product of vertex operators of tachyons. Its OPE with 
$\tau > \tau'$ is 
\beeqno
e^{i p X} (\tau) e^{i q X}(\tau') &\sim&
 \langle e^{i p X} (\tau) e^{i q X}(\tau')\rangle   \\   
&=&(\tau-\tau')^{2 \alpha'G^{\alpha \beta} p_{\alpha} q_{\beta}}
e^{^\frac{1}{2} i \theta^{\alpha \beta} p_{\alpha} q_{\beta}}e^{i(p+q)X}.
\eneqno
If the factor $(\tau-\tau')^{2 \alpha'G^{\alpha \beta} p_{\alpha} q_{\beta}}$ 
can be ignored, this is the star product;
\beeq
e^{i p X} (\tau) e^{i q X}(\tau') \sim  e^{i p X} * e^{i q X}(\tau).  \label{a}
\eneq
For this reason we consider $\alpha' \to 0$ limit in which only massless
fields are dominant. In the Seiberg-Witten limit
\beeqno
\alpha' &\sim& \epsilon^{\frac{1}{2}} \to 0   \\
g_{\alpha \beta}&\sim& \epsilon    \to   0
\eneqno
$G$ and $\theta$ are
\beeq
G^{\alpha \beta}&\to& -\frac{1}{(2 \pi \alpha')^2 }
(\frac{1}{B}g\frac{1}{B})^{\alpha \beta}      \nonumber  \\
G_{\alpha \beta}&\to& -(2 \pi \alpha')^2 (Bg^{-1}B)_{\alpha \beta}    \nonumber  \\
\theta ^{\alpha \beta} &\to&\left( \frac{1}{B} \right)^{\alpha \beta}.  \label{10-9}
\eneq
(\ref{10-9}) is consistent with the result in section 3.7.
And the propagator at boundary (\ref{10-6}) becomes
\beeqno
\langle X^{\alpha}(\tau) X^{\beta}(0) \rangle 
=\frac{i}{2} \theta^{\alpha \beta}\epsilon(\tau).
\eneqno
We can generalize the equation (\ref{a}) to a product of any functions. By 
performing formal power series in the limit we have
\beeqno
:f(x(\tau))::g(x(0)):\sim :(f*g)(x(0)):.
\eneqno 
And it is important to notice that at boundary on the worldsheet in the zero slope 
limit, the OPE of a derivative of $x$ and $x$ does not equivalent to the star product, 
\beeqno
:\part_{\tau}X^{\a}(\tau)::X^{\b}(0):\sim 0 \hsp1 ({\rm for} \ \tau>0)
\eneqno
since the propagator is proportional to the sign function $\epsilon(\tau)$. More generally 
we can say that the OPE of any polynomial in derivatives of $x$ does not equivalent 
to the star product.  
\par
Next, we would like to see that the low energy effective action of a D-brane 
with a point splitting 
reguralization used is described by a non-commutative gauge theory\cite{seiberg}.   
We consider a background gauge field with a coupling term;
\beeqno
-i \int d \tau A_{\alpha}(X)\partial_{\tau}X^{\alpha}.
\eneqno 
At a classical level this term is invariant under the gauge transformation 
(\ref{8-5});
\beeqno
\delta \int  d \tau A_{\alpha}(X)\partial_{\tau}X^{\alpha}
=\int d \tau \partial_{\tau} \lambda
\eneqno 
which is a total derivative. However at a quantum level the coupling term is in 
the argument of an exponential in the partition function, and we have to perform 
a regularization procedure. As well known the Pauli-Villars regularization 
preserves the U(1) gauge invariance, and the low energy effective action has 
the $U(1)$ gauge invariance \cite{fradkineffective,fradkinelectro,tseytlin}. Let us 
consider a point splitting reguralization. 
A relevant factor 
\beeq
\exp[-i\int  A]    \label{10-7}
\eneq
in the partition function 
is transformed by the gauge transformation.
Expanding the exponential in powers of $A$ and $\lambda$, the first order term in A is 
\beeqno
-\int  d \tau A_{\alpha}(X)\partial_{\tau}X^{\alpha} 
\int d \tau' \partial_{\tau'}\lambda.
\eneqno
The point splitting regularization is a procedure that we cut out the region 
$|\tau-\tau'|<\delta$ ($\delta$ is infinitesimal) in the integral.
\beeqno
-\int  d \tau A_{\alpha}(X)\partial_{\tau}X^{\alpha} 
(\int^{\infty}_{\tau+\delta}+\int^{\tau-\delta}_{-\infty})
 d \tau \partial_{\tau'}\lambda.  
\eneqno
We explicitly write the normal ordering symbol which is omitted in the above
\beeqno
-\int  d \tau :A_{\alpha}(X)\partial_{\tau}X^{\alpha}: 
:[\lambda(X(\tau^-))-\lambda(X(\tau^+))]:\sim \int  d \tau (A_{\alpha}*\lambda-
\lambda*A_{\alpha})\partial_{\tau}X^{\alpha}  .
\eneqno
Here the OPE has been performed. Hence the point splitting regularization 
violates the gauge invariance (\ref{8-5}). 
We see that (\ref{10-7}) is invariant under the non-commutative gauge transformation
(\ref{6-4}). n-th order terms in A of 
\beeqno
\exp[-i\int(A+\partial \lambda +i[\lambda,A]_*)]
\eneqno
are
\beeqno
&&\frac{i^{n+1}}{n!} 
\int A(X(t_1))\cdots A(X(t_n))\partial_t \lambda (X(t))   \\
&&+\frac{i^{n+1}}{(n-1)!}\int  
A(X(t_1))\cdots A(X(t_{n-1}))(\lambda(X(t_n))*A-A*\lambda(X(t_n))),
\eneqno
after performing the point splitting reguralization and the OPE the first term 
becomes
\beeqno 
&&\frac{i^{n+1}}{n!} 
\int :A(X(t_1))\cdots A(X(t_n))::\partial_t \lambda (X(t)):  \\
&&\hsp1 \sim \frac{i^{n+1}}{n!} \int \sum_j 
A(X(t_1))\cdots A(X(t_{j-1}))A(X(t_{j+1}))\cdots A(X(t_n))  \\
&&\hspace{4cm}\times (A*\lambda (X(t_j))-\lambda *A(X(t_j)))
\eneqno
each term in the last line are identical;
\beeqno
=\frac{i^{n+1}}{(n-1)!} A(X(t_1))\cdots  A(X(t_{n-1}))
(A*\lambda (X(t_n))-\lambda *A(X(t_n))).
\eneqno
Therefore the first and second terms are canceled out with each other. 
The theory has the non-commutative gauge invariance instead of the ordinary gauge 
symmetry. The extension of this consequence to 
the NSR superstring is also discussed in\cite{seiberg}.


\chapter{The Seiberg-Witten Map}
\section{The Seiberg-Witten Map}
In chapter 3 we have seen that a Dp-brane with a constant gauge field strength 
is equivalent to a configuration of $\infty$ D(p-2)-branes. The gauge field on the 
Dp-brane is an ordinary U(1) gauge field. On the other hand if we consider the 
worldvolume as that of D(p-2)-branes the worldvolume theory is the non-commutative 
gauge theory. In the 
last chapter it was observed the symmetry of the theory is depend on the regularization. 
If we use the Pauli-Villars regularization ordinary gauge invariance is preserved. 
In contrast to this the point splitting regularization violates the gauge 
invariance and we have the non-commutative gauge theory. Usually two S-matrix which 
are obtained by using two different regularizations are related by a transformation.
From this observation, there must be a map from ordinary to non-commutative gauge
fields\cite{seiberg,asakawa}, which is called as the Seiberg-Witten map. 
 We construct the transformation so that the map may 
preserve the gauge equivalence\cite{seiberg}. An ordinary gauge field 
A\footnote{The gauge theory is of arbitrary rank. We will deal with a simple case, 
namely, U(1) gauge theory later.  } is gauge 
equivalent to $A+\delta_{\lambda}A$, and an non-commutative gauge field $\hat{A}$ 
is gauge equivalent to $\hata+\delta_{\hatlam}\hata$ where $\delta_{\lambda}$ and 
$\delta_{\hatlam}$ are ordinary  and non-commutative gauge transformations with 
parameters $\lambda$ and 
$\hatlam$. 
\beeqno
A           & \longrightarrow& \hspace{3cm} A+\delta_{\lambda}A     \\
\downarrow                   && \hspace{35mm}  \downarrow            \\
\hata(A)    & \longrightarrow&  \hata(A)+\delta_{\hatlam}\hata(A) 
= \hata (A+\delta_{\lambda} A)
\eneqno
From the requirement that the Seiberg Witten map preserves the gauge equivalence we have 
\beeq
\hata(A)+\delta_{\hatlam}\hata(A) = \hata (A+\delta_{\lambda} A).   \label{10-10}
\eneq
We would like solutions of this equation up to first order in $\theta$. 
We set as
\beeqno
\hata&=&A+A'(A)  \\
\hatlam&=&\lambda +\lambda'(\lambda,A)
\eneqno
where $A'(A)$ and $\lambda'(\lambda,A)$ are of first order in $\theta$ and vanish when 
$\theta=0$. 
(\ref{10-10}) is written as
\beeq
A'_i(A+\delta_{\lambda}A)-A'_i-\partial_i \lambda' -i[\lambda',A_i]-i[\lambda,A'_i]
=-\frac{1}{2}\theta^{kl}(\partial_k 
\lambda \partial_l A_i +\partial_l A_i \partial_k \lambda).    \label{10-11}
\eneq
It is noticed that 
\beeq
\hata_i&=& A_i -\frac{1}{4}\theta^{kl} \{A_k,\partial_lA_i+F_{li} \}  
+\calo(\theta^2)                                         \label{10-12} \\
\hatlam&=&\lambda+\frac{1}{4}\theta^{ij}\{\partial_i \lambda ,A_j \} 
+\calo(\theta^2) \nonumber
\eneq
are solutions of (\ref{10-11}). Here $\{ \ ,\  \}$ is an anti-commutator. However 
it is easily noticed that the solutions have an ambiguity since the two 
functions $\hata_i$ and $\hatlam$ are derived form one equation 
(\ref{10-11})\cite{asakawa}. There are other source of ambiguities\cite{asakawa}, 
but we do not discuss this in this paper.  Although the calculation is tedious, the 
relation between ordinary and non-commutative field strength can be obtained by 
using (\ref{10-12}) as
\beeqno
\hat{F}_{ij}=F_{ij}+\frac{1}{4}\theta^{kl}(2 \{F_{ik},F_{jl}\}-\{A_k,D_lF_{ij}
+\partial_lF_{ij }\})+\calo(\theta^2).
\eneqno
\par
Similarly we can construct a transformation from a non-commutative gauge field 
on a space with parameter $\theta$ to a non-commutative gauge field with 
$\theta+\delta \theta$. This case with $\theta=0$ corresponds to the above case. 
Demanding that the map preserves the gauge equivalence relation
\beeqno
\hata(\theta) 
& \longrightarrow& \hspace{5cm} \hata+\delta_{\hatlam}\hata(\theta)     \\
\downarrow                   && \hspace{55mm}  \downarrow            \\
\hata(\theta+\delta \theta)   
& \longrightarrow&  
\hata(\theta+\delta \theta)+\delta_{\hatlam}\hata(\theta+\delta \theta) 
= \hata (\hata(\theta)+\delta_{\hatlam} \hata(\theta))
\eneqno   
gives 
\beeq
\hata(\theta+\delta \theta)+\delta_{\hatlam}\hata(\theta+\delta \theta) 
= \hata (\hata(\theta)+\delta_{\hatlam} \hata(\theta)).   \label{10-13}
\eneq
The Moyal product on non-commutative space with non-commutativity 
parameter $\theta+\delta \theta$ is 
\beeqno
\exp[i\frac{\theta^{ij} +\delta \theta^{ij}}{2}
\partial_i ^\alpha \partial_j^\beta]
f(x+ \alpha)g(x+\beta)|_{\alpha=\beta=0}
&=&\frac{i}{2}\delta \theta^{ij}\frac{\partial f}{\partial x^i}*
\frac{\partial g}{\partial x^j}    \\
&=&\delta \theta^{ij}\frac{\partial}{\partial \theta ^{ij}}(f*g)
\eneqno
where (\ref{6-5}) is used. Let
\beeqno
\hata(\theta+\delta \theta)&\equiv&\hata(\theta)+\delta \hata(\theta)  \\
\hatlam(\theta+\delta \theta)&\equiv&\hatlam(\theta) +\delta \hatlam(\theta).
\eneqno
Using this,
\beeqno
&&\delta \hata_i (\hata(\theta) +\delta_{\lambda}\hata(\theta))
-\delta \hata_i (\theta)-\partial_i \hatlam (\theta)
+i[\delta \hatlam(\theta),\hata_i (\theta)]_* 
+i[\hatlam(\theta),\delta \hata_i (\theta)]_*   \\
&&\hsp1 =i \delta \theta ^{ij} \frac{\partial}{\partial \theta^{ij}}
[\hatlam(\theta),\hata_i (\theta)]_* 
\eneqno
whose solutions are given by 
\beeq
\delta \hata_i(\theta) &=&-\frac{1}{4}\delta \theta^{kl}[\hata_k*(\partial_l \hata_i
+\hat{F}_{li})+(\partial_l \hata_i+\hatf_{li})*\hata_k]    \nonumber   \\
\delta \hatlam(\theta)&=&\frac{1}{4}\delta \theta^{kl}(\partial_k \hatlam *\hata_l
+\hata_l* \partial_k \hatlam)              \nonumber   \\
\delta \hat{F}_{ij}&=&\frac{1}{4}\d \theta^{kl}[2 \hat{F}_{ik}*\hat{F}_{jl}
+2 \hat{F}_{jl}*\hat{F}_{ik}-\hata_k *(\hat{D}_l\hat{F}_{ij}
+\partial_l \hat{F}_{ij})  \nonumber \\
&&\hsp1 -(\hat{D}_l\hat {F}_{ij}  +\partial_l \hat{F}_{ij})*\hata_k ].  \label{10-23}
\eneq
If we would like a map from $\theta=0$ to a finite $\theta$, we need to integrate
the above. For simplicity we deal with a U(1) gauge theory with a constant field 
strength. For this case  (\ref{10-23}) is reduced to 
\beeqno
\delta \hat{F}_{ij}=-\delta \theta^{kl}\hat{F}_{ik} \hat{F}_{lj},
\eneqno
we rewrite this in  the Lorentz indices omitted form as
\beeq
\delta \hat{F}=-\hat{F}\delta \theta \hat{F}.   \label{10-14}
\eneq 
The solution of the differential equation with condition $\hat{F}(\theta=0)=F$ is
\beeq
\int^{\hat{F}}_F d \hat{F}\frac{1}{\hat {F}^2}&=&-\int^{\theta}_0 d \theta 
                                             \nonumber  \\
\hat{F}&=&\frac{1}{1+F\theta} F.                     \label{10-15}
\eneq
As a check a variation of this result is (\ref{10-14}).
The ordinary field strength in terms of the non-commutative field strength is 
written as
\beeqno
F=\hat{F}\frac{1}{1-\theta \hat{F}} .
\eneqno
In $\alpha'\to0$ limit $\theta=B^{-1}$. When $F+B=0$, it is noticed from 
(\ref{10-15}) that we can not use the 
non-commutative description of the gauge theory. The criterion of 
whether we can use the non-commutative gauge theory or not is gauge 
invariant.
\par
The effective theory of a D-brane for slowly varying fields is the Dirac-Born-Infeld 
Lagrangian \cite{tseytlin};
\beeqno
\call(F) =\frac{1}{g_s (2 \pi)^p(\alpha')^{\frac{p+1}{2}}}
\sqrt{\det(g+2 \pi \alpha'(F+B))}.
\eneqno 
If we use the point splitting regularization, a field strength is $\hatf$ 
and products of fields are the star products. Then the action\cite{leedbi} is 
\beeqno
\hat{\call}(\hatf) =\frac{1}{G_s (2 \pi)^p(\alpha')^{\frac{p+1}{2}}}
\sqrt{\det(G+2 \pi \alpha'\hatf)}_*.
\eneqno
From (\ref{10-15}) $F=0$ corresponds to $\hatf=0$. Then the constant part of 
the two Lagrangian have to be equivalent.
Then we have
\beeqno
G_s=g_s \left(\frac{\det G}{\det (g+ \pial B)}\right) ^{\frac{1}{2}}
\eneqno
which becomes in the zero slope limit 
\beeq
G_s=g_s\det(\pial Bg^{-1})^{\frac{1}{2}}.   \label{10-25}
\eneq
\par
In the above we have seen that two different regularization scheme give 
commutative and non-commutative gauge theories. The point splitting regularization 
of the two dimensional theory on the string worldsheet with boundary gives the 
non-commutative gauge theory 
with field strength $\hatf$ and the Moyal product. The Lagrangian is a function of 
$\hatf$ and NS B field appear implicitly in the Moyal product and the open string 
metric $G_{ij}$. (In the zero slope limit 
the non-commutativity parameter $\theta$ is equivalent to $B^{-1}$.) On the other 
hand if we use the Pauli-Villars regularization, in the DBI action the gauge field 
and NS B field appear in the form $F+B$ and the product is ordinary one. It is 
natural to guess there are other regularizations which correspond to other 
non-commutative descriptions of the gauge theory. Seiberg and Witten\cite{seiberg} 
suggested that the B dependence in the DBI action appear in $\theta$ and some field 
$\Phi$ which is in the form of $\hatf +\Phi$. We see their argument below. However 
they did not give a proof for 
this. In order to prove this conjecture we have to look for another regularization 
which gives a non-commutative theory. Further investigation is 
needed\footnote{The $\Phi$ dependence is discussed in \cite{seibergmatrix,andreev} 
which are informed by O. Andreev}.
\par
From the definition of the metric $G$ and non-commutativity $\theta$, they guessed 
relations between $\Phi$ and other variables are  
\beeq
\frac{1}{G+\pial \Phi}&=&-\frac{\theta}{\pial}+\frac{1}{g+\pial B} \label{10-16} \\
G_s&=&
g_s \left(\frac{\det (G+ \pial \Phi)}{\det (g+ \pial B)}\right) ^{\frac{1}{2}}.
                                                       \label{10-17}
\eneq
The second equation is from the suggestion for the $\Phi$ dependence in the DBI 
action. We have some consistency checks. When $\theta=0$ (\ref{10-16}) and 
(\ref{10-17}) are consistent with $G=g$, $G_s=g_s$ and $\Phi=B$. Next when $\Phi=0$ 
(\ref{10-16}) and (\ref{10-17}) are reduced to the representation of the theory 
which has been observed in the last chapter which is derived from the point splitting 
regularization. We consider $\alpha' \to 0$ limit where 
\beeqno
g&=&\epsilon g^{(0)} +\calo(\epsilon^2)  \\
B&=&B^{(0)}+\epsilon B^{(1)}+\calo(\epsilon^2)  \\
\alpha' &=&\epsilon^{\frac{1}{2}}
\eneqno
so that $G$ and $\Phi$ may be 0 th order in $\epsilon$. 
In the limit from (\ref{10-16})
\beeq
\theta&=&\frac{1}{B^{(0)}}  \label{10-20}   \\
G&=&-\frac{(\pial)^2}{\epsilon}B^{(0)}\frac{1}{g^{(0)}}B^{(0)}  \label{10-18}   \\
\Phi&=&-B^{(0)}+\frac{(\pial)^2}{\epsilon}B^{(0)}\frac{1}{g^{(0)}}B^{(1)}  
\frac{1}{g^{(0)}}B^{(0)}.                     \label{10-19}  \\
G_s&=&g_s \det \left( \frac{\pial}{\epsilon}B^{(0)}\frac{1}{g^{(0)}}\right).
                                              \label{10-21}  
\eneq
(\ref{10-20}), (\ref{10-18}) and (\ref{10-21}) are coincide with the zero slope 
limit of the results of the point splitting regularization. 
\par
As a final check we see that there is no inconsistency for the combination 
$\hatf+\Phi$ in the DBI Lagrangian. In the zero slope limit the Lagrangian is 
$\tr (\hatf+ \Phi)^2$. It 
is easily understood that the existence of the $\Phi$ does not affect to a physical 
observable. The Lagrangian is 
\beeqno   
\tr (\hatf+ \Phi)^2=\tr (\hatf^2 +2 \hatf \Phi +\Phi^2).
\eneqno
The second and third terms are a total derivative and a constant term respectively 
which do not contribute to S-matrix. Then we can neglect the $\Phi$ dependent terms.  
\par
So far we have seen regularization dependence of the theory with the NS B 
field fixed. Here we would like to vary the B field with the point splitting 
regularization used. In the commutative description there is a symmetry with respect 
to the B field, namely 
\beeqno
B_{ij} \to B_{ij}+\partial_i \Lambda_j-\partial_j \Lambda_i.
\eneqno
In the non-commutative description is there a counterpart of this symmetry ? 
\par
For simplicity we deal with $\theta$ whose rank is equivalent to the 
dimensionality  of that of the D-brane so that $\theta$ may be invertible. The 
covariant derivative of the gauge theory is  defined by 
\beeqno
D_i=\partial_i-iA_i
\eneqno
on the non-commutative space $[x^i,x^j]=\theta^{ij}$.
We rewrite it as 
\beeq
D_i=\partial'_i-iC_i.    \label{10-22}
\eneq
where
\beeqno
\partial'_i&=&\partial_i+i B_{ij}x^j   \\
C_i&=& A_i+B_{ij}x^j.
\eneqno
$\partial'$ commute with $x^i$; $[\partial'_i,x_j]=0$.
For the sake of the unusual definition (\ref{10-22}), the field strength takes the  
simple form 
\beeqno
\hatf_{ij}
=B_{ij}-i[C_i,C_j].
\eneqno 
Introduce vierbeins $e^{a}_i$ and $E^a _i$ with respect to the metric $g$ and $G$  
respectively ($g_{ij}=\sum_a e^{a}_i e^{a}_j,\ G_{ij}=\sum_a E^{a}_i E^{a}_j$). 
$a$ is an index of the local Lorentz frame.  We vary $\theta$ with $g_{ij}$ (in 
other words $e^a _i$) and $C_a$ fixed ($C_a$ is defined by $C_i=E^a _i C_a$). 
Using the form of the metric $G$ in $\alpha \to 0$ limit  
\beeqno
G^{il}&=&-(\pial)^{-2}\theta^{ij}g_{jk}\theta^{kl}
\eneqno
gives that the vierbein for the metric $G$ is $E^a _i= \pial B_{ij}e^j _a$.
We would like to show  that   the Lagrangian
\beeq
G^{ik} G^{jl}\tr (\hatf_{ij}-\theta_{ij}^{-1})*(\hatf_{kl}-\theta_{kl}^{-1})
                                                 \label{10-24}
\eneq
is background independent. The problem is reduced to the work to prove 
\beeqno
Q^{il}=-i \theta^{ij}[C_j,C_k]\theta^{kl}
\eneqno 
is background independent which is from $\theta ^{ij}C_i$ is independent;
$\theta^{ij}C_i=-(\pial)C_a e^{ja}$.
Hence the Lagrangian is background independent. 
\par
In the commutative description only $\L$ gauge invariant function is $B+F$. This 
is coincide with the above result.  In the case of U(1) gauge theory 
with a constant field strength, using (\ref{10-15}) gives $Q$ in terms of the 
ordinary field strength;
\beeqno
Q
=-\frac{1}{B+F}
\eneqno
where we have used;
\beeqno
\hatf-B=-B\frac{1}{B+F}B.
\eneqno
In the both descriptions of the DBI theory $Q$ is a background independent 
value. We can define the transformation rule of a coupling constant so that 
a measure of the action will be background independent. 
\section{Dirac-Born-Infeld Action} 
We have three different descriptions of the  gauge theory. In this section we would 
like to see that these descriptions are equivalent to each other. The low energy 
effective theory 
of a D-brane in a slowly varying field approximation is the Dirac-Born-Infeld 
Lagrangian. As we have seen in the last section we can describe a D-brane by 
the Lagrangian 
\beeqno
\hat{\call} =\frac{1}{G_s(2 \pi)^p (\alpha')^{\frac{p+1}{2}}} 
\sqrt{\det (G+ \pial (\hatf+\Phi))}_* 
\eneqno
which is derived by using the point splitting regularization. In the Lagrangian 
products of functions are the Moyal products. Here what we would like 
to do is to prove the non-commutative DBI Lagrangian is equivalent to the ordinary DBI 
action up to total derivative and derivative terms of field strength of a gauge field. 
The DBI Lagrangian is valid for the case in which derivatives of the field strength is 
negligible. Then we replace the Moyal product to ordinary one. For simplicity 
we set $\pial=1$. The process of the proof is only to vary $\theta$ in the 
non-commutative DBI Lagrangian with $g,B$ and $g_s$ fixed. We will see its result is 
total derivative terms and $\calo(\t^2)$. We take a variation of (\ref{10-16}) with 
$g$, $B$ and $g_s$ fixed; 
\beeqno
\delta G+\delta \Phi=(G+\Phi)\delta \theta (G+\Phi).
\eneqno
We take a variation of (\ref{10-17});
\beeqno
\delta G_s
&=&\frac{G_s}{2} \tr (G+\Phi)\delta \theta \\
&=&\frac{G_s}{2} \tr (\Phi \delta \theta)
\eneqno
where the fact that $G$ is a symmetric matrix, $\Phi$ and $\theta$ are 
anti-symmetric matrices is used. In the present situation (\ref{10-23}) becomes 
\beeqno
\delta \hatf_{ij}(\theta )=\delta \theta^{kl}\left[ \hatf_{ik}\hatf_{jl}
-\frac{1}{2} \hata_k (\partial_l \hatf_{ij}+\hatd_l \hatf_{ij})\right] 
+\calo (\partial \hatf).
\eneqno
We take a variation of the Lagrangian with respect to $\theta$
\beeq
&&\delta\left[ \frac{1}{G_s}\sqrt{\det (G+  \hatf+\Phi)}\right]   \nonumber   \\
&& \hsp1 = \frac{1}{G_s}\sqrt{\det (G+  \hatf+\Phi)}\left[ 
-\frac{\delta G_s}{G_s} +\frac{1}{2}\tr \frac{1}{G+  \hatf+\Phi}
(\delta G+  \delta \hatf+ \delta \Phi) \right]          \nonumber  \\
&& \hsp1 =\fra12 \frac{1}{G_s}\sqrt{\det (G+  \hatf+\Phi)} [
-\tr \delta \theta (G+\Phi )                
+ \tr \frac{1}{G+  \hatf+\Phi}
(G+\Phi)\delta \theta (G+\Phi)                 \nonumber  \\
&&\hspace{2cm}+ \left(\frac{1}{G+  \hatf+\Phi} \right)_{ji}
\delta \theta^{kl}( \hatf_{ik}\hatf_{jl}  
-\fra12 \hata_k (\partial_l \hatf_{ij}+\hatd_l \hatf_{ij})) ]
+\calo (\partial \hatf).                         \label{10-27}
\eneq
Here we need to write  some calculations explicitly. 
\beeqno
\tr \frac{1}{G+  \hatf+\Phi}(G+\Phi)\delta \theta (G+\Phi)
=\tr \delta \theta (G+\Phi) 
-\tr \frac{1}{G+\hatf+\Phi}\hatf \delta \theta (G+\Phi)  \\
\partial_l \det (G+  \hatf+\Phi)^{\frac12}=\fra12 \det (G+  \hatf+\Phi)^{\fra12} 
\left(\frac{1}{G+  \hatf+\Phi} \right)_{ji}\partial_l \hatf_{ij}    \label{10-26}   \\
\hatd_l \det (G+  \hatf+\Phi)^{\frac12}=\frac12 \det (G+  \hatf+\Phi)^{\frac12} 
\left(\frac{1}{G+  \hatf+\Phi} \right)_{ji}\hatd_l \hatf_{ij}.
\eneqno
The last term in (\ref{10-27}) is rewritten as
\beeqno
&&\fra12 \frac{1}{G_s}\sqrt{\det (G+  \hatf+\Phi)}
 \left(\frac{1}{G+  \hatf+\Phi} \right)_{ji}
\delta \theta^{kl}(-\fra12) \hata_k (\partial_l \hatf_{ij}+\hatd_l \hatf_{ij})  \\
&&\hsp1 =\fra12 \frac{1}{G_s}\delta \theta^{kl} \hatf_{lk}\sqrt{\det (G+  \hatf+\Phi)}  
+({\rm total \ derivative})
\eneqno
Hence (\ref{10-27}) becomes 
\beeqno
&=&\fra12\frac{1}{G_s}\det (G+\hatf +\Phi)^{\fra12} [-\tr \frac{1}{G+\hatf +\Phi}
\hatf \delta \theta (G+\Phi)-\tr\frac{1}{G+\hatf +\Phi}
\hatf \delta \theta\hatf+\tr \delta \theta \hatf ] \\
&&+\calo(\partial \hatf)+{\rm total \ derivative}  \\
&=&\calo(\partial \hatf)+{\rm total \ derivative}. 
\eneqno
Then we have the conclusion that two action whose non-commutativity parameters are 
different by $\delta \theta$ ($\delta \theta$ is infinitesimal) will give a same 
physical S-matrix. Furthermore infinite chain of the variation would give an 
equivalence of DBI Lagrangians with non-commutativity parameters whose difference is 
finite. Hence the effective theory of a D-brane  has infinitely many equivalent 
descriptions. It will be important to see whether properties of D-branes (T-duality 
etc.), which are satisfied in the ordinary commutative description, are depend on 
descriptions or not. This is a future problem.  
\par
In constructing the effective theory of a D-brane derivative terms of fields were 
omitted\cite{fradkineffective}. We need to find the derivative corrections and a 
systematic way to construct these terms. 
It was pointed out by Okawa and Terashima\cite{okawadbi,okawa} that derivative 
corrections to the DBI Lagrangian can be constructed  
by the equivalence of ordinary gauge theory and non-commutative gauge theory.

\chapter{Canonical Quantization of an Open String with NS B Field}
\section{Dirac Formalism}
In the canonical formalism a  canonical momentum for a given Lagrangian $L(q,\dot{q})$ 
is defined by
\beeqno
p_i =\frac{\partial L}{\partial \dot{q}^i}.  
\eneqno
The Hessian matrix is given by 
\beeqno 
W_{ij}\equiv \frac{\partial^2 L}{\partial \dot{q}^i \partial \dot{q}^j}.
\eneqno
If $\det W_{ij}=0$, the Lagrangian is called singular. For a singular Lagrangian 
we can not write the $\dot{q}^i$ in terms of the canonical momentum which is need if 
we construct a Hamiltonian from the Lagrangian. For example, in Yang-Mills gauge 
theory, the time component of a momentum $p^0$ is zero. Hence this system is 
singular. Dirac developed the method how 
to deal with the singular system \cite{dirac,nakanishi}. In this section we will 
review this in a formal way.
\par
In the Lagrangian and Hamiltonian formalism, independent variables are 
$(q,\dot{q})$ and $(q,p)$ respectively. Because of the singularity, the Jacobian 
$\frac{\partial p_i}{\partial \dot{q}^j}$ of 
the Legendre transformation is zero. This means that $p_i$ are 
not independent of $q^i$ and $\dot{q}^i$ which are related by primary constraints
\beeqno
\Phi_a ^{(0)}(q,p)=0  .
\eneqno    
The total Hamiltonian is 
\beeqno
H_T=H+\lambda_a \Phi_a
\eneqno
where $\lambda_a$ are Lagrange multipliers. The equation of motion for a variable 
$g(q,p)$ is 
\beeqno
\dot{g}(q,p) =\{ g(q,p), H_T \}_p.
\eneqno
The primary constraints should be invariant under the time evolution;
\beeq
\dot{\Phi}_a ^{(0)}(q,p)&=&\{ \Phi_a ^{(0)},H_T \}_p         \nonumber   \\
&=&\{\Phi_a^{(0)},H \}_p+\lambda_b \{ \Phi_a^{(0)},\Phi_b^{(0)}\} =0. 
                                                      \label{11-1}
\eneq
In this step there are some possibilities. 
\begin{itemize}
\item
(\ref{11-1}) holds identically

\item
$\{ \Phi_a^{(0)},\Phi_b^{(0)}\}$ vanishes and this gives  new constraints
\beeqno
\Phi_a ^{(1)} \equiv \{ \Phi_a ^{(0)},H \}_p .
\eneqno

\item
$\{ \Phi_a^{(0)},\Phi_b^{(0)}\}$ does not vanish and we obtain $\lambda_a$ for 
which (\ref{11-1}) is satisfied.
\end{itemize}
Next similarly we should do the same procedure for $\Phi^{(1)}$.
And repeat this procedure until it does not give new constraints. The set of 
constraints $\Phi^{(1)},\Phi^{(2)},\cdots$ are called secondary constraints. 
In quantum field theory, there is the  possibility that this process does not have 
end, in other words, there are infinite constraints. However for almost cases, the 
number of secondary constraints is finite. In the next section we will consider the 
system of an open string coupled to the NS B field. We will regard boundary 
conditions of the open string as primary constraints. The constraints will give 
infinitely many secondary constraints.
\par
There is another classification of constraints. If Poisson brackets of a constraint 
with all other constraints (both primary and secondary) vanish, the constraint is 
called first class. Other constraints are second class.
\par
In a non-singular system the process of the quantization is equivalent to the 
replacement of the Poisson bracket with the commutator
\beeqno
i\{q,p\}_p  \to [q,p].
\eneqno
Dirac suggested that when all constraints in theory is second class, the 
quantization  is given by the replacement 
\beeqno      
i\{q,p\}_D  \to [q,p]
\eneqno
where 
\beeqno
\{A,B\}_D \equiv \{A,B\}_p-\{A,\Phi_M \}(C^{-1})^{MN}\{\Phi_N,B\}_p
\eneqno
is Dirac bracket ($C^{MN}\equiv \{\Phi^M,\Phi^N \}_p$). 
\par
If there are also first class constraints the matrix $C^{MN}$ does not have an 
inverse matrix. The first class constrains correspond to generators of 
transformations. For instance gauge theory includes first class constraints. In 
this case we have to fix the symmetries which correspond to the first class 
constraints. We choose the gauge fixing condition so 
that all of the constraints will become second class. 
\section{Dirac Quantization of an Open String}
The action of an open string with the gauge and NS B fields is given by
\beeqno
S=-\frac{1}{4 \p \a '} \int d \s ^2 [\part_a X^{\alpha} 
\partial^aX_{\beta}+ \calf_{\a \b} \epsilon^{ab}\partial_a X^{\alpha}
\partial_bX^{\beta}]
\eneqno  
where $\calf=F-B$ and Minkowski signature is used. We fix the $\L$ gauge symmetry 
as $\calf=-B$. 
The boundary condition is given by 
\beeqno
\partial_{\sigma}X_{\alpha}+B_{\alpha \beta}\partial_{\tau}X^{\b}=0 
            \hsp1 {\rm at} \ \sigma=0,\pi
\eneqno
and canonical momentum is 
\beeqno
P^{\alpha}=\frac{1}{\pial} (\partial_{\tau}X^{\alpha}
+B^{\alpha  \b}\partial_{\sigma}X_{\b}).
\eneqno    
This action is seen not to  have primary constraints. Then as usual we 
would like to give equal $\tau$ canonical commutators as
\beeq
\left[X^{\a}(\s),P^{\b}(\s')\right]&=&i\d ^{\a \b}\d (\s-\s') \nonumber  \\ 
\left[X^{\a}(\s),X^{\b}(\s')\right] &=& 0   \label{11-2}       \\
\left[P^{\a}(\s),P^{\b}(\s')\right]&=&0.      \nonumber 
\eneq
However this does not coincide with the boundary condition. We can easily understand 
this. In terms of the canonical momentum the boundary conditions are
\beeqno
0
=M_{\a \b}\partial_{\s}X^{\b}+\pial B_{\a \b}P^{\b}
\eneqno
where $M\equiv\e-B^2$. From this we have
\beeqno
-\pial B_{\a \b}\left[P^{\b},P^{\g}\right]&=&M_{\a \b}
\part_{\s}\left[X^{\b},P^{\g} \right]  \\
-\pial B_{\a \b}\left[P^{\b},X^{\g}\right]&=&M_{\a \b}
\part_{\s}\left[X^{\b},X^{\g} \right].
\eneqno
If (\ref{11-2}) is satisfied the commutator 
$\left[X^{\a}(\s),P^{\b}(\s')\right]$ should vanish. Hence the ordinary commutators 
are not true in this system. Then we need a way to determine the commutators. 
\par
In the case with no background fields, usually we use the symplectic form 
\beeqno
\O =\int d{\s} dP_{\a}dX^{\a}.
\eneqno
Using the mode expansion (\ref{2-11}), we can find the symplectic form for 
the modes which is $\tau$ independent. On the other hand if there is the NS B field 
the symplectic form for modes is $\tau$ dependent. Then it is suggested 
in\cite{chunoncom} that instead of $\O$, we should use its time average 
\beeqno
<\O>=\lim_{T \to \infty} \frac{1}{2 T} \int^T _{-T} \O d \tau
\eneqno
as the symplectic form which is $\tau$ independent\footnote{However in reality the 
symplectic form $\O$ does not have $\tau$ dependence even if there is the NS B field 
background. The unnecessary procedure in \cite{chunoncom} has been corrected in 
\cite{chuneutral}. This is informed by C.-S. Chu.}. This gives the commutator 
which is consistent with the boundary condition.
\par
There is another method which will be reviewed in this section. 
We regard the boundary conditions as primary 
constraints \cite{ardalan,chu,sheikh,leecano,leeopen,fayyazuddin,kim}. It will be 
found that the constraints are second class. The Dirac bracket of secondary constrains 
with a canonical variable identically vanish. Then the commutator which is 
constructed by this method is manifestly consistent with the boundary 
conditions.
\par 
We introduce primary constraints;
\beeqno
\Phi^{\a}(0)=\Phi^{\a}(\p)=0 
\eneqno
where
\beeqno
\Phi^{\a}(\s)\equiv \pial B^{\a}_{\ \b}P^{\b}+\partial_{\s}X^{\b}M^{\a}_{\ \b} \hsp1
M^{\a}_{\ \b}=(\e-B^2)^{\a}_{\ \b}.
\eneqno
The procedure which we have seen in the previous section gives secondary constraints 
\beeqno
\phi^{(1\a n)}\equiv \part_{\s}^{2n}\Phi^{\a}(\s)=0,
\hsp1 \phi^{(2\a n)}\equiv\part_{\s}^{2n+1}P^{\a}(\s)=0  \hsp1 (n=0,1,\cdots)
\eneqno
at $\s=0,\p$. We have infinite constraints. Here we would like to see that these 
constraints are second class. 
We set as
\beeqno
C^{(i \a n)(j \b m )}=\{ \phi^{(i \a n)},\phi^{(j \b m)} \}_p
\eneqno
where $i=1,2$. Components of the $C^{(i \a 0)(j \b 0 )}$ are
\beeqno
\{ \Phi^{ \a}(\s),\Phi^{\b }(\s') \}_p
&=&-\pial (BM)^{\a \b}[\part_{\s}\d (\s-\s')+\part_{\s'}\d (\s-\s')]   \\
\{ \Phi^{ \a}(\s),\part_{\s '}P^{\b }(\s') \}_p&=& M^{\a \b}\part_{\s}\part_{\s'}
\d(\s-\s')                 \\
\{ \part_{\s}P^{ \a}(\s),\part_{\s '}P^{\b }(\s') \}_p&=& 0.
\eneqno
For $n,m\neq 0$, we have
\beeqno
C^{(i \a n)(j \b m )}=\part^{2n}_{\s}\part^{2m}_{\s'}C^{(i \a 0)(j \b 0 )}.
\eneqno
The constraints are defined only at $\s=0,\p$. Then we treat indices $\s,\s',\cdots$ 
as discrete variables below. Its inverse matrix is formally written as
\beeqno
(C^{-1})_{(i \a n)(j \b m )}(\s'',\s''')=\left(
\begin{array}{cc}
0 & -(M^{-1})_{\a \b}R_{nm}(\s'',\s''')   \\
(M^{-1})_{\a \b}R_{nm}(\s'',\s''')   & \pial(BM^{-1})_{\a \b}S_{nm}(\s'',\s''')   \\
\end{array}
\right)
\eneqno
where matrices $R$ and $S$ satisfy
\beeq
\sum_{m \s''}\part^{2n+1}_{\s}\part^{2m+1}_{\s''}\d(\s-\s'')R_{mp}(\s'',\s''')&=&
\d^n_{p}\d_{\s \s'''}                   \label{11-3}                 \\
\sum_{m \s''}\part^{2n}_{\s}\part^{2m}_{\s''}
[\part_{\s}\d(\s-\s'')+\part_{\s''}\d(\s-\s'')]
R_{mp}(\s'',\s''')&=&\sum_{m \s''}\part^{2n+1}_{\s}\part^{2m+1}_{\s''}\d(\s-\s'')
S_{mp}(\s'',\s''').         \nonumber   \\
          \label{11-4}
\eneq
We would like to find Dirac brackets. It is trivial that 
\beeqno
\{P^{\a}(\s),P^{\b}(\s')\}_D=0.
\eneqno 
Next we would like to see $\{X^{\a}(\s),X^{\b}(\s')\}_D$. From the definition of the 
Dirac bracket, 
\beeq
&&\{X^{\a}(\s),X^{\b}(\s')\}_D    \nonumber  \\
&&\hspace{1.5cm}=-\pial \sum_{n m \s'' \s'''}[
B^{\g \a}(M^{-1})_{\g}^{\b}\part^{2n}_{\s''}\d(\s -\s'')R_{nm}(\s'',\s''')
\part^{2m+1}_{\s'''}
\d(\s'''-\s')   \nonumber  \\
&&\hspace{2cm} -\d^{\a \g}\part^{2n+1}_{\s''} \d(\s-\s'')(M^{-1})_{\g \d}
R_{nm}(\s'',\s''')
\part^{2m}_{\s'''}\d(\s'''-\s')B^{\b \d}  \nonumber \\
&&\hspace{2cm}+\d^{\a \g}\part^{2n+1}_{\s''} 
\d(\s-\s'')(BM^{-1})_{\g \d}S_{nm}(\s'',\s''')
\part^{2m+1}_{\s'''}\d(\s'''-\s')\d^{\b\d}]       \label{11-6} 
\eneq
If $\s ,\s'\neq 0, \pi$, $\{X^{\a}(\s),X^{\b}(\s')\}_D=0$. We would like to see
the value of the Dirac bracket for the case $\s ,\s'= 0$. (\ref{11-4}) for $n=0$ and 
multiplying $\sum_{p}\sum_{\s'''}\part^{2p+1}_{\s'''}\d(\s'-\s''')$ and integrate 
over $\s$ we have 
\beeqno
&&\int d\s \sum_{m p \s'' \s'''}\part^{2m}_{\s''}
[\part_{\s}\d(\s-\s'')+\part_{\s''}\d(\s-\s'')]
R_{mp}(\s'',\s''')\part^{2p+1}_{\s'''}\d(\s'-\s''')    \\
&&\hsp1 =\int d \s \sum_{m p \s'' \s'''}\part_{\s}\part^{2m+1}_{\s''}\d(\s-\s'')
S_{mp}(\s'',\s''') \part^{2p+1}_{\s'''}\d (\s'-\s''').
\eneqno     
Using
\beeqno
\int d \s \part_{\s ''}\d(\s''-\s)&=&\part_{\s''}\int d\s \d(\s''-\s)  \\
&=&0
\eneqno
gives
\beeq
&&\left. \sum_{p \s'' \s'''}\part^{2m}_{\s''}
\d(\s-\s'')R_{mp}(\s'',\s''')\part^{2p+1}_{\s'''}\d(\s'-\s''')
\right|^{\s=\p}_{\s=0}          \nonumber     \\
&&\hsp1= \left. \sum_{p \s'' \s''}\part^{2m+1}_{\s''}\d(\s-\s'')
S_{mp}(\s'',\s''') \part^{2p+1}_{\s'''}\d (\s'-\s''') \right|^{\s=\p}_{\s=0}. 
                                      \label{11-7}
\eneq
By virtue of the formula (\ref{11-7}), the Dirac bracket is simplified in the 
following combination of the brackets;  
\beeq
&&\{X^{\a}(\p)-X^{\a}(0),X^{\b}(\s')\}_D             \nonumber   \\
&&\hsp1=\{X^{\a}(\p),X^{\b}(\s')\}_D-\{X^{\a}(0),X^{\b}(\s')\}_D   \nonumber   \\  
&& \hsp1 =-\pial(BM^{-1})^{\a \b}
\left. \sum_{ \s'' \s'''}\part^{2n+1}_{\s''}
\d(\s-\s'')R_{nm}(\s'',\s''')\part^{2m}_{\s'''}\d(\s'''-\s')
\right|^{\s=\p}_{\s=0}  \nonumber          \\
               \label{11-5}
\eneq
where (\ref{11-6}) has been used. Using (\ref{11-3}) for $n=0$ multiplying it with 
$\sum_{p}\sum_{\s'''}\part^{2p}_{\s'''}\d(\s'-\s''')$ and integrate over $\s$, we 
have
\beeq
&& \int d \s \sum_{m p \s '' \s'''}\part_{\s}\part^{2m+1}_{\s''}
\d(\s-\s'')R_{mp}(\s'',\s''')\part^{2p}_{\s'''}\d(\s'-\s''')  \nonumber  \\
&& \hsp1 =\int d\s
\sum_{p \s'''} \d^0_{p}\d_{\s, \s'''}  \part^{2p}_{\s'''}\d(\s'-\s''')  \nonumber  \\
&&\hsp1=\sum_{\s'''}\d (\s'-\s''')         \nonumber  \\
&&\hsp1= \d(\s')+\d (\s'-\p).           \label{11-8}
\eneq
At a same time there are not both $X(0)$ and $X(\p)$ in a constraint in this 
theory. Then we have
\beeqno
\{X^{\a}(0),X^{\b}(\p)\}_D=0.
\eneqno
From (\ref{11-5}) and (\ref{11-8}) we have
\beeqno
\{X^{\a}(0),X^{\b}(0)\}_D&=&\pial (M^{-1}B)^{\a \b}\d (0)   \\
\{X^{\a}(\p),X^{\b}(\p)\}_D&=&-\pial (M^{-1}B)^{\a \b}\d (0)  .
\eneqno
In this method of calculation we can not determine the explicit form of 
$\{X^{\a}(\s),P^{\b}(\s')\}_D$ at boundaries. Of course 
$\{X^{\a}(\s),P^{\b}(\s')\}_D$ takes usual form at $\s,\s'\neq 0,\p$. 
\section{Generalization to the NSR Superstring}
So far we have seen that the effect of the NS B field  to the commutation relations of 
string coordinates. In this section we extend this to the fermionic variables in 
the NSR formalism \cite{chu,leeopen}. In the conformal gauge we have to add the terms;
\beeq
S=\frac{i}{2 \p} \int d^2 \s \bar{\psi}^{\mu}\r ^{a} \part_{a} \psi_{\mu}
-\frac{i}{4 \p\a'} \int d^2 \s B_{\a \b}\bar{\psi}^{\a}\epsilon^{ab} \r_{a} \part_{b}
 \psi_{\b}              \label{11-9}
\eneq
where
\beeqno
\psi=\left(
\begin{array}{c}
\psi    \\
\tilde{\psi}     \\
\end{array}
\right)
\eneqno
is a majorana spinor and 
\beeqno
\r^0 =\left(
\begin{array}{cc}
0   &   -i   \\
i   &   0    \\
\end{array}
\right)  \hsp1    
\r^1 =\left(
\begin{array}{cc}
0   &   i   \\
i   &   0    \\
\end{array}
\right)
\eneqno
are 2 dimensional gamma matrices. 
Here we would like to find the boundary conditions for the fermionic variables 
in a manifestly supersymmetric way \cite{chen}. Using a superfield $\Phi$ the NSR 
superstring action is given by 
\beeq
S=\frac{1}{2\p} \int dz d\bar{z} d\t d\bar{\t}(g_{\mu \nu}+\pial B_{\mu \nu})
\bar{D}\Phi^{\mu}(\mathbf{z},\bar{\mathbf{z}})D\Phi_{\mu}(\mathbf{z},\bar{\mathbf{z}})
                                  \label{11-10}
\eneq
where $\mathbf{z}=(z,\t)$ and $\bar{\mathbf{z}}=(\bar{z},\bar{\t})$ are coordinates 
of the worldsheet superspace and
\beeqno
D&=&\frac{\part}{\part \t} +\t \frac{\part}{\part z}    \\
\bar{D}&=&\frac{\part}{\part \bar{\t}} +\bar{\t} \frac{\part}{\part \bar{z}}
\eneqno
are the super covariant derivatives. The superfield is written in terms of the 
component fields as
\beeqno
\Phi^{\mu}(\mathbf{z},\bar{\mathbf{z}})=\sqrt{\frac{2}{\a'}} X^{\mu} (z,\bar{z}) 
+i \t \psi^{\mu}(z,\bar{z}) +i \bar{\t} \tilde{\psi}^{\mu}(z,\bar{z})
+i \t \bar{\t} F^{\mu} (z,\bar{z})
\eneqno
We can eliminate the auxiliary field $F^{\mu}$ by the equation of motion. Then the 
action is rewritten as
\beeqno
S=\frac{1}{2\p} \int dz d\bar{z} d\t d \bar{\t}(g_{\mu \nu}+\pial B_{\mu \nu})
\left( \frac{2}{\a'} \bar{\part}X^{\mu}\part X^{\nu}-\bar{\part}\psi^{\mu}\psi^{\nu}
+\tilde{\psi}^{\mu} \part \tilde{\psi}^{\nu} \right)
\eneqno
which is same with (\ref{11-9}). 
We set $\tilde{g}_{\mu \nu}=g_{\mu \nu}+\pial B_{\mu \nu}$ and take a variation of 
the action (\ref{11-10});
\beeqno
\d S=\frac{1}{2\p} \int dz d\bar{z} d\t d \bar{\t}\tilde{g}_{\mu \nu} 
(\bar{D}\delta \Phi^{\mu}(\mathbf{z},\bar{\mathbf{z}})
D\Phi^{\nu}(\mathbf{z},\bar{\mathbf{z}})
+\bar{D}\Phi^{\mu}(\mathbf{z},\bar{\mathbf{z}})D\delta \Phi^{\nu}
(\mathbf{z},\bar{\mathbf{z}})).
\eneqno
$\frac{\part}{\part \t}$ term does not contribute to the boundary condition. 
Then we have
\beeqno
\int d\t d \bar{\t}\tilde{g}_{\mu \nu} [\bar{\t} (\d \Phi ^{\mu} D \Phi^{\nu})
+\bar{D}\Phi^{\mu} \d \Phi ^{\nu} \t]=0. 
\eneqno
Along longitudinal directions of a D-brane 
\beeqno
\left. g_{\a \b } (\part-\bar{\part})X^{\b}
+\pial B_{\a \b } (\part+\bar{\part})X^{\b} \right|_{\s=0,\p}=0,    
\eneqno
for NS fermion
\beeqno
\left. g_{\a \b } (\psi^{\b}-\tilde{\psi}^{\b})
+\pial B_{\a \b } (\psi^{\b}+\tilde{\psi}^{\b}) \right|_{\s=0,\p}=0,
\eneqno
and for R fermion
\beeqno
\left. g_{\a \b } (\psi^{\b}-\tilde{\psi}^{\b})
+\pial B_{\a \b } (\psi^{\b}+\tilde{\psi}^{\b}) \right|_{\s=0}=0,    \\
\left. g_{\a \b } (\psi^{\b}+\tilde{\psi}^{\b})
+\pial B_{\a \b } (\psi^{\b}-\tilde{\psi}^{\b}) \right|_{\s=\p}=0.
\eneqno
\par
Although we can get the commutator for the fermionic variables by repeat the process 
in the previous section, instead we find the commutators by using the supersymmetry 
on the worldsheet explicitly. The symmetry is defined by 
\beeqno
\d_{\epsilon} X^{\mu}&=&\bar{\epsilon}\psi^{\mu}      \\
\d_{\epsilon}\psi^{\mu}&=& -i \r^{a}\part_a X^{\mu} \epsilon.
\eneqno
Since $X$ and $\psi$ are not mixed in the boundary conditions, 
\beeqno
\{\psi^{\mu},X^{\nu} \}_{D}=0
\eneqno
is easily understood. The equation must be invariant under the supersymmetry 
transformation 
\beeq
0&=&\d  \{\psi^{\mu},X^{\nu} \}_{D}   \nonumber   \\
&=&\left(
\begin{array}{c}
\{ -\tilde{\epsilon}\part_0 X^{\mu}+\tilde{\epsilon}\part_1X^{\mu},X^{\nu}\}_D   \\
\{ \epsilon \part_0 X^{\mu}+\epsilon\part_1X^{\mu},X^{\nu}\}_D   \\
\end{array}
\right)  +\left(
\begin{array}{c}
\{ \psi^{\mu} (\s),i \tilde{\epsilon}\psi^{\nu} -i\epsilon \tilde{\psi}^{\nu} \}_D   \\
\{ \tilde{\psi}^{\mu} (\s),i \tilde{\epsilon}\psi^{\nu} 
-i\epsilon\tilde{\psi}^{\nu}\}_D   \\
\end{array}
\right) .             \label{super}
\eneq
Because of the Dirichlet conditions the parameters are related by 
\beeqno
\epsilon =\l \tilde{\epsilon}   \hsp1 \l = \pm 1
\eneqno
for which the Dirichlet conditions are preserved. We are aware the Dirac brackets of 
X, then by using (\ref{super}) we  have the Dirac brackets of $\psi$.  
For example for the NS fermion at boundaries the commutators are given by 
\beeqno
\{\psi^{\a}, \psi^{\b} \}_D 
&=&\p \a ' i \eta^{\a \b} \tilde{\d}(\s-\s')   \\
\{\tilde{\psi}^{\a}, \tilde{\psi}^{\b} \}_D 
&=&\p \a ' i \eta^{\a \b} \tilde{\d}(\s-\s')   \\
\eneqno
where
\beeqno
\tilde{\d}(\s-\s')=\d(\s-\s') -\sum_{\s'',\s'''}\part^{2m+1}_{\s''} \d(\s-\s'')
R_{mk}(\s'',\s''')\part^{2k+1}_{\s'''} \d(\s'''-\s').
\eneqno
For $\s,\s'\neq 0,\p$ these are reduced to the ordinary commutators. Although we can 
not obtain explicit form of the Dirac bracket (the $R_{mk}(\s'',\s''')$ is an 
unknown function) these seems to be B independent. If this is true, the B 
independence of commutators of fermions is 
common in NSR and GS superstring theories \cite {chusuper}.


\chapter{Summary and Remarks}
In this paper we have seen relations between commutative and non-commutative 
spacetime. In chapter 3 we have reviewed in the framework of the boundary state 
formalism that a Dp-brane with a constant field strength of gauge field is 
equivalent to $\infty$ D($p-2$)-branes. The worldvolume theories of the Dp-brane and 
$\infty$ D(p-2)-branes are DBI theory on ordinary and non-commutative space, 
respectively. 
From the result, we have found that we can represent coordinates on the D-brane 
by commutative and also non-commutative one. In chapter 4 it has been reviewed that 
using point splitting regularization violates $U(1)$ gauge symmetry, and make the 
theory non-commutative $U(1)$ gauge invariant. In chapter 5 we could construct a 
transformation which connects an ordinary gauge field and a non-commutative gauge 
field which was first advocated by Seiberg and Witten. It is need to see whether the 
properties (T-duality etc.) which hold in commutative representation hold in 
non-commutative representation or not.
\par  
In chapter 6 we have found commutation relations which coincide with boundary 
conditions in the framework of the operator formalism. In the case that there is a 
constant background NS B field, we treat the boundary conditions (which is mixed 
type) as primary constraints, and we have non-commutative coordinates at endpoints 
of an open string. Because there are infinite secondary constraints, it is difficult 
to find commutators for a non-constant B field. We would like a method to deal with 
the case. 
\par
We obtained the commutators in the bosonic and NSR string theories. The 
case of Green-Schwarz string with supergravity background \cite{grisaru} is studied 
in \cite{chusuper}. 
\par
Last of all, we would like to comment on a relation between results in these 
chapters. It is common in chapter 4 and 6 that the source of the non-commutativity 
of coordinates is the NS B field. The difference is the methods. The CFT and the 
operator formalism gives the same result. The connection between non-commutativities 
from dimension of D-brane and reguralization is discussed in \cite{katononcom} in the 
case of compactified space.

\end{document}